\documentclass[preprint,showpacs,preprintnumbers,amsmath,amssymb,aps]{revtex4-1}
\usepackage{graphicx,dcolumn,bm}% Include figure files
\usepackage{amssymb,amstext,amsmath}

\usepackage{graphicx}% Include figure files
\usepackage{dcolumn}% Align table columns on decimal point
\usepackage{bm}% bold math
\usepackage{amssymb}
\usepackage{multirow}
\usepackage{bigstrut}
\usepackage{makecell,rotating}

\usepackage{mathrsfs}
\usepackage{booktabs}
\usepackage{threeparttable}
\usepackage{multirow}
\usepackage{subfigure}
\usepackage{epsfig}
\usepackage{threeparttable}
\usepackage{chngpage}
\usepackage{float}

\usepackage{amstext}% bold math
\usepackage{amsmath}% bold math
\usepackage[right]{lineno}
\usepackage{leftidx}

\usepackage{hyperref}
\hypersetup{
    bookmarks=true,
    colorlinks=false, %set true if you want colored links
    linktoc=all,     %set to all if you want both sections and subsections linked
    linkcolor=red,  %choose some color if you want links to stand out
}

\begin{document}

\title{Evolutionary multiplayer games on graphs with edge diversity}
\author{Qi Su$^{1,2}$,  Lei Zhou$^{1}$, and Long Wang$^{1, }$}
\email{longwang@pku.edu.cn}
\affiliation{
$^{1}$Center for Systems and Control, College of Engineering, Peking University, Beijing 100871, China \\
$^{2}$Center for Polymer Studies, Department of Physics, Boston University, Boston, Massachusetts, United States of America
}

\begin{abstract}
Evolutionary game dynamics in structured populations has been extensively explored in past decades.
However, most previous studies assume that payoffs of individuals are fully determined by the strategic behaviors of interacting parties and social ties between them only serve as the indicator of the existence of interactions.
This assumption neglects important information carried by inter-personal social ties such as genetic similarity, geographic proximity, and social
closeness, which may crucially affect the outcome of interactions.
To model these situations, we present a framework of evolutionary multiplayer games on graphs with edge diversity, where different types of edges describe diverse social ties.
Strategic behaviors together with social ties determine the resulting payoffs of interactants.
Under weak selection, we provide a general formula to predict the success of one behavior over the other.
We apply this formula to various examples which cannot be dealt with using previous models, including the division of labor and relationship- or edge-dependent games.
We find that labor division facilitates collective cooperation by decomposing a many-player game into several games of smaller sizes.
The evolutionary process based on relationship-dependent games can be approximated by interactions under a transformed and unified game.
Our work stresses the importance of social ties and provides effective methods to reduce the calculating complexity in analyzing the evolution of realistic systems.
\end{abstract}
\maketitle

\section{Introduction}
Understanding the emergence and persistence of cooperation in the population of egoists is an enduring challenge that has inspired a myriad of studies from biology to sociology \cite{1995-MaynardSmith-p-}.
Evolutionary game theory has been widely employed to investigate this cooperation conundrum at different levels of living systems \cite{1982-MaynardSmith-p-}.
Typically, social dilemmas are depicted by two-player two-strategy games where each player can choose either to cooperate or to defect \cite{2002-Macy-p7229-7236}.
In these games, mutual cooperation brings each player a reward $R$ while mutual defection a punishment $P$;
when a cooperator encounters a defector, the cooperator obtains a sucker's payoff $S$ and the defector gets the temptation $T$.
Different rankings of payoff entries $R,S,T,P$ represent different social dilemmas \cite{2002-Macy-p7229-7236}.
Despite the simplicity of this representation, in the real world, many interactions occur beyond the dyadic scenarios and often involve more than two individuals.
For examples, in a S. cerevisiae population, a cooperative yeast produces an enzyme to hydrolyze sucrose into monosaccharides while the most of them diffuse away and are exploited by nearby yeasts \cite{2009-Gore-p253-253}
(see Ref.~\cite{2003-Rainey-p72-72,2004-Griffin-p1024-1024} for more examples in microbes and Ref.~\cite{1968-Hardin-p1243-1248,2006-Milinski-p3994-3998} in human societies).
Interactions in these examples are better modeled by multiplayer games \cite{2010-Gokhale-p5500-5504}.
Generally, multiplayer games cannot be represented by a collection of two-player games \cite{2016-McAvoy-p203-238} whereas the latter can always be regarded as the simplest case of the former \cite{2010-Gokhale-p5500-5504}, making the study of multiplayer games of great importance for the evolution of cooperation \cite{1997-Broom-p931-952,2018-Wu-p1-15}.
One particular example is the threshold public goods game \cite{2009-Pacheco-p315-321}.
It captures the strategic interactions of individuals when the provision of public goods needs a threshold surpassed.
Such a threshold can be a minimum amount of funding for building national defense, a minimum height of a dam for securing the public safety, etc \cite{2006-Milinski-p3994-3998}.
In this game, each individual has two options---to contribute an amount of investment to the goods pool or not to contribute.
The benefit is provided only when the total investment exceeds a threshold \cite{2009-Pacheco-p315-321}.

Recent advance in exploring interaction patterns of living agents shows that populations often exhibit structural characteristics, which expands our research interests in evolutionary dynamics from traditional well-mixed to structured populations \cite{1992-Nowak-p826-826,2004-Hauert-p643-646,2007-Szabo-p97-216,2009-Fu-p46707-46707,2013-Perc-p-,2015-Du-p8014-8014,2015-Zhou-p60006-60006,2015-Pena-p122-136,2016-Li-p22407-22407,2017-Allen-p227-227,2017-Su-p103023-103023}.
Graphs serve as a good tool to model such a system, where vertices of graphs represent individuals and edges
specify one's interaction and dispersal neighborhoods.
In the case of weak selection where individuals' payoffs obtained from games slightly affect their fitness or reproductive rates,
evolutionary outcomes on graphs, especially the conditions for one strategy to be favored over the other, can be tackled analytically.
For example, Tarnita \emph{et al.} derive a simple condition to predict the evolutionary outcome for two-player two-strategy games \cite{2009-Tarnita-p570-581}.
This condition relies on all the payoff entries $R,S,T,P$ and one ``structure coefficient".
As shown in their work, the structure coefficient summarizes all the effects of a population structure on the condition for the success of strategies and it is independent of payoff entries.
Due to the generality of the above results, calculating structure coefficients provides a convenient way to quantify the effect of population structures on the evolutionary outcome \cite{2009-Nathanson-p1-7,2011-Tarnita-p2334-2337,2014-Debarre-p3409-3409,2016-Zhang-p1-16,2016-McAvoy-p203-238,2017-Allen-p227-227,2013-Wu-p-}.
Nonetheless, the closed-form expressions of the structure coefficients are often hard to calculate under multiplayer games, even in the simplest well-mixed populations \cite{2010-Gokhale-p5500-5504}.
This becomes even more challenging when the population structure is taken into account.
Even so, there are still a few seminal work about evolutionary multiplayer games on graphs \cite{2014-Li-p5536-5536,2016-Pena-p1-15,2012-Broom-p70-80,2016-Pena-p-,2013-Wu-p-,2016-McAvoy-p203-238}.
For example, Pe$\tilde{\text{n}}$a \emph{et al.} derive the structure coefficients for evolutionary multiplayer games on finite ring graphs and infinite regular graphs \cite{2016-Pena-p1-15}.
Based on competition between territorial animals, Broom \emph{et al.} develop a new modelling framework to investigate collective interactions, which is capable and flexible to compare and analyze various spatial structures \cite{2012-Broom-p70-80}.
McAvoy \emph{et al.} study when a multiplayer game can be broken down into a sequence of interactions with fewer individuals and show that a simple population structure can greatly complicate the reduction \cite{2016-McAvoy-p203-238}.

Prior studies about games on graphs usually assume that social ties between individuals only indicate the presence of interactions \cite{2011-Veelen-p116-128,2012-Broom-p70-80,2013-Wu-p-,2014-Li-p5536-5536,2015-Zhou-p60006-60006,2016-Pena-p1-15,2016-Pena-p-,2016-McAvoy-p203-238, 2016-Su-p103007-103007}.
The other relevant information associated with social ties, such as the genetic and physical relationships between interactants, is often ignored.
In such cases, individuals' strategic behaviors are the only determinant of the outcome of an interaction.
Typically, in two-player interactions, if two distinct individuals take the same strategy, their common opponent obtains the same payoff when encountering each of them separately \cite{1992-Nowak-p826-826,2004-Hauert-p643-646}.
When engaging in group interactions, one's payoff relies on the number of opposing cooperators but is independent of which one is the cooperator \cite{2016-Pena-p1-15,2014-Li-p5536-5536}.
Indeed, this assumption significantly reduces the calculation complexity and thus makes it possible for many well-known results \cite{2006-Ohtsuki-p502-502,2006-Nowak-p1560-1563}.
However, recent studies show that overlooking the information of social ties could make theoretical predictions deviate greatly from empirical observations \cite{2004-Pastor-Satorras-p-, 2007-Onnela-p7332-7336,2015-Pastor-Satorras-p925-979,2009-Wuchty-p15099-15100}.
For example, people possess strong and weak social ties, such as intimate interpersonal relationships with relatives and tenuous relationships with acquaintance \cite{1973-Granovetter-p1360-1380,2009-Wuchty-p15099-15100};
failing to account for the tie strengths leads to a globally accelerated information diffusion and a remarkably distinct diffusion direction from that in actual networks \cite{2007-Onnela-p7332-7336,2015-Pastor-Satorras-p925-979}.
In well-mixed populations, when distinct frequencies of interactions between pairs are considered, altruistic traits can flourish whereas neglecting such information on social ties leads to the extinction of altruism \cite{2017-Allen-p227-227}.
Here, the second example clearly conveys that the information associated with social ties can affect the evolution of a certain behavioral trait (strategy) in a nontrivial way.
Besides, we offer two other representative cases.
In the example of the division of labor in colonies of eusocial insects and human societies,
the production of collective benefits needs different individuals to cooperatively perform different subtasks \cite{2014-Wright-p9533-9537,2001-Franks-p635-642,2005-Kay-p165-174,1986-Franks-p425-429}.
When many individuals assigned one subtask cooperate, cooperation from an individual assigned another subtask is more crucial to the colony productivity than cooperation from individuals assigned the same subtask.
%Whether the required types of individuals are present is crucial to survival of all the individuals involved.
The other situation is that the payoff structure of an interaction may be relationship-dependent \cite{2014-Maciejewski-p117-128}.
It means that an individual may concurrently play various types of games with its neighbors, depending on the social tie they are connected with \cite{2000-Cressman-p67-81,2006-Hashimoto-p669-675}.
For instance, individuals can play coordinations games (or even harmony games) with its friends and prisoner's dilemma with strangers.

To better understand the role of social ties in the evolution of strategic behaviors, we present a comprehensive framework of evolutionary multiplayer games on graphs with edge diversity.
Each type of edges describes one kind of relationship between two connected individuals, such as having the same or different task skills \cite{2001-Franks-p635-642,2005-Kay-p165-174,1986-Franks-p425-429,1992-Stander-p445-454}, owning close or distinct consanguinity or geographical distance and so on.
We investigate both finite and infinite regular graphs with $n$ types of edges.
We provide a simple condition to predict when natural selection favors one strategic behavior over the other.
The condition is validated by Monte Carlo simulations.
Applying it to the case of division of labor where cooperation from individuals performing different subtasks is required for producing benefits (see the example of army ants retrieving prey items \cite{2001-Franks-p635-642}),
we find labor division significantly lowers the barrier to establish cooperative society.
Then we explore the scenario where each individual simultaneously participates in many multiplayer games and these games can differ in payoff entries or metaphors.
We find evolutionary dynamics for such diverse interactions can be approximated by an evolutionary process with a unified payoff structure.
This result provides us insights into simplifying complex and diverse interactions in real-world systems as simple and unified interactions in theoretical calculations.
Our work also covers the evolutionary games on weighted graphs (see the example of bacterium Escherichia coli \cite{2013-Allen-p1169-1169}).
Intriguingly, in our framework, strong edges do not act as a promoter of cooperation.

\section{Models}
\begin{figure*}[!h]
\centering
\includegraphics[width=0.8\textwidth]{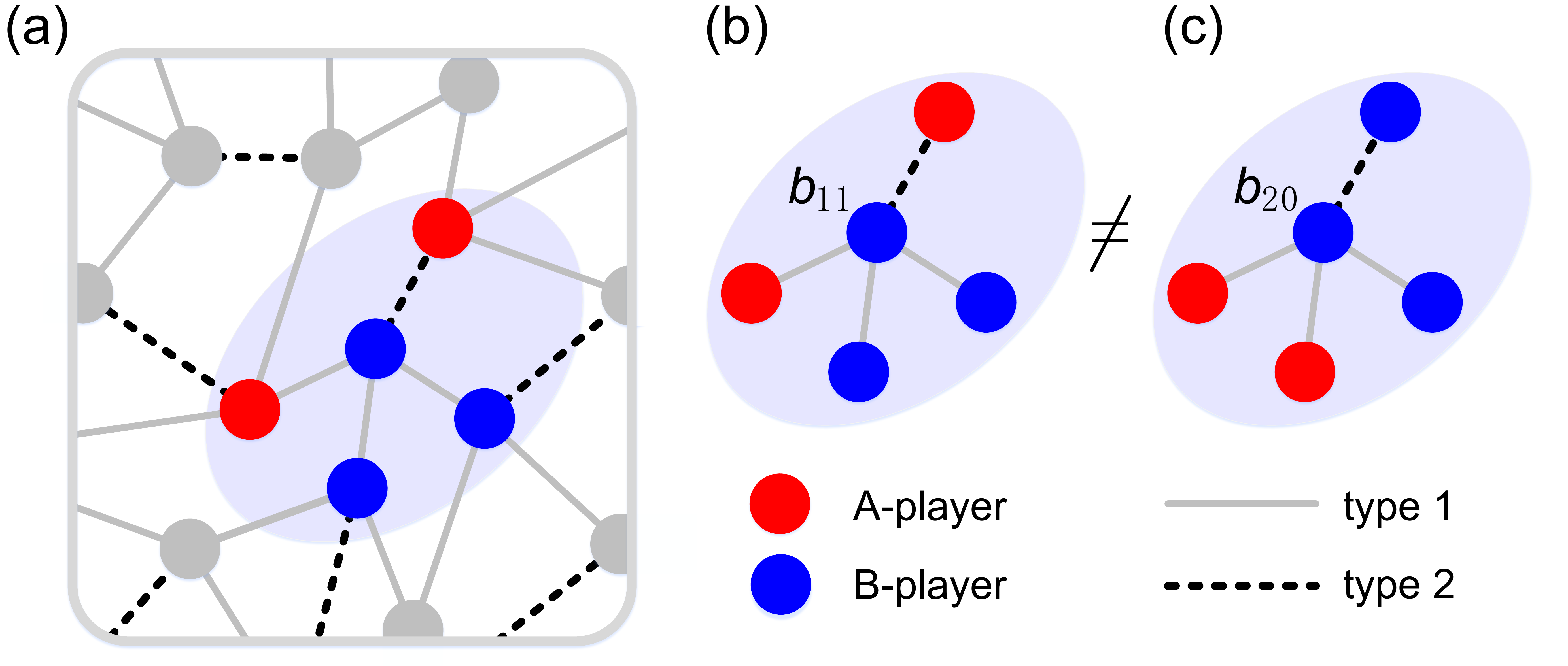}
\caption{\label{Fig 1} \textbf{Illustration of evolutionary multiplayer games on graphs with two types of edges}.
(a) Each node is linked to $4$ other nodes ($k=4$) by two types of edges, one marked by solid line ($g_1=3$) and the other marked by dashed line ($g_2=1$).
Each node is occupied by an individual, either $\text{A}$- (red circle) or $\text{B}$-player (blue circle).
One's payoff is determined by the strategies of its own and all individuals occupying neighboring nodes.
For example, in the highlighted area, all individuals altogether determine the payoff of the centered individual.
(b) Interacting with an A- and two B-players linked by edges of type $1$, and an A-player linked by an edge of type $2$, the centered B-player gains a payoff $b_{11}$.
(c) The centered B-player obtains a payoff $b_{20}$ when interacting with two A- and a B-player linked by edges of type $1$, and a B-player linked by an edge of type $2$.
Note that $b_{11}$ differs from $b_{20}$ although the total number of neighboring A-players is the same.
}
\end{figure*}

Here we briefly introduce the model of evolutionary multiplayer games on graphs with edge diversity.
We first consider the stochastic evolutionary dynamics on a graph-structured population with a finite size $N$ and later investigate the dynamics in infinite populations.
Each individual occupies a node of a random regular graph with degree $k$.
Each node is linked to $k$ other nodes by $n$ types of edges $(1\leq n \leq k)$, where the number of type $i$ is $g_i$, i.e., $\sum_{i=1}^ng_i=k$.
Note that after determined randomly, this graph is fixed during the process of evolution.
Each individual chooses a strategy between A and B.
In each generation, every individual obtains a payoff by interacting with $k$ adjacent individuals in a single game, analogous to the setting of spatial multiplayer game in prior studies \cite{2002-Szabo-p118101-118101,2016-Pena-p1-15}.
If there are $s_i$ opposing A-players and $g_i-s_i$ opposing B-players among interaction partners linked by edges of type $i$ ($1\le i\le n$), the focal A-player gets a payoff $a_{s_1s_2\cdots s_n}$ whereas the focal B-player gets a payoff $b_{s_1s_2\cdots s_n}$.
Fig \ref{Fig 1} illustrates an example of the spatial structure and Table \ref{Table 1} presents its payoff structure for $n=2$.
Our model can recover the traditional setting by taking $n=1$.

\begin{table}
%\begin{adjustwidth}{-2.25in}{0in} % Comment out/remove adjustwidth environment if table fits in text column.
\caption{\label{Table 1}
\textbf{Payoffs for A- and B-players in multiplayer games with two types of interaction partners.} } \centering
\begin{tabular}{p{3.4cm}<{\centering}| p{1.4cm}<{\centering} p{1.4cm}<{\centering} p{1.4cm}<{\centering} p{0.3cm}<{\centering} p{1.4cm}<{\centering}  p{0.8cm}<{\centering} p{1.8cm}<{\centering} p{1.8cm}<{\centering}  p{1.4cm}<{\centering}}
\hline
\hline
Opposing A-players (Type 1, Type 2) & $(0,0)$ & $(0,1)$ & $(1,0)$ & $\cdots$ & $(s_1,s_2)$ & $\cdots$ & $(g_1-1,g_2)$ & $(g_1,g_2-1)$ & $(g_1,g_2)$\\
\hline
Payoff to A & $a_{00}$ & $a_{01}$ & $a_{10}$ & $\cdots$ & $a_{s_1s_2}$ &$\cdots$ &$a_{(g_1-1)g_2}$ & $a_{g_1(g_2-1)}$ & $a_{g_1g_2}$ \\
\hline
Payoff to B & $b_{00}$ & $b_{01}$ & $b_{10}$ & $\cdots$ & $b_{s_1s_2}$ &$\cdots$ &$b_{(g_1-1)g_2}$ & $b_{g_1(g_2-1)}$ &
$b_{g_1g_2}$ \\
\hline
\hline
\end{tabular}
\begin{flushleft} Taking social closeness for example, $s_1$ is the number of opposing $A-$players among $g_1$ partners with close social relationships and $s_2$ the number of opposing $A-$players among $g_2$ partners with distant social relationships.
\end{flushleft}
%\end{adjustwidth}
\end{table}

After the interaction, individual $i$'s payoff $\pi_i$ is transformed to its reproductive rate or fitness by $F_i=1-\omega+\omega\pi_i$.
$\omega$ represents the intensity of selection, i.e., the extent to which the payoff from games influences the reproductive success.
Here we consider the weak selection ($\omega\ll 1$).
The population evolves according to the death-birth process \cite{2006-Ohtsuki-p502-502}.
Concretely, a random individual such as $i$ is selected to die.
After that, $i$'s neighbors compete to replace the vacancy with probability proportional to their reproductive rate.
This update rule can also be translated into a rule for behavior imitation \cite{2007-Szabo-p97-216,2017-Allen-p227-227}.
For example, a random individual $i$ resolves to update its strategy, and it adopts neighbor $j$'s strategy proportionally to $j$'s fitness, i.e., with probability $F_j/\sum_{l\in \Omega_i}F_l$, where $\Omega_i$ is the set of $i$'s neighbors.
In this paper, we view the updating process as a kind of behavior imitation (other update rules can be analyzed analogously).

\section{Results}
\subsection{A general condition to predict the success of one strategic behavior.}
In finite populations, the fixation probability is a well-established measure to quantify the evolutionary success of different traits or strategies \cite{2004-Nowak-p646-646}.
The fixation probability $\rho_{\text{A}}$ denotes the probability that a single A-player starting in a random position propagates and takes over the whole population of B-players.
Analogously, $\rho_{\text{B}}$ is the probability that a single B-player starting in a random position propagates and takes over the whole population of A-players.
Natural selection favors strategy A over B if
\begin{linenomath}\begin{align}
\rho_{\text{A}} > \rho_{\text{B}}. \nonumber
\end{align}\end{linenomath}
Using weak selection, in large random regular graphs with $n$ edge types ($k\ge 3$ and $0\le g_i \le k$), we obtain the condition under which A-players are selected over B-players (see S1 Text, Section 1), given by
\begin{linenomath}\begin{align} \label{sigma}
\sum_{s_1=0}^{g_1}\sum_{s_2=0}^{g_2}\cdots\sum_{s_n=0}^{g_n}\sigma_{s_1s_2\cdots s_n}\left(a_{s_1s_2\cdots s_n}-b_{(g_1-s_1)(g_2-s_2)\cdots (g_n-s_n)}\right)>0
\end{align}\end{linenomath}
where $\sigma_{s_1s_2\cdots s_n}$ ($0 \le s_1 \le g_1$, $0 \le s_2 \le g_2$, $\cdots$, $0 \le s_n \le g_n$) is the structure coefficient that relies on population structures and update rules but is independent of payoff values $a_{s_1s_2\cdots s_n}$ and $b_{s_1s_2\cdots s_n}$.
There are totally $\Pi_{i=1}^{n}(g_i+1)$ structure coefficients for Eq~(\ref{sigma}).
All structure coefficients here are positive and we can eliminate an extra structure coefficient through dividing the sigma rule [see Eq~(\ref{sigma})] by any one of them.
$\sigma_{s_1s_2\cdots s_n}$ can be approximated by
\begin{linenomath}\begin{align}
\sigma_{s_1s_2\cdots s_n}=
&\frac{(k-2)^{(k-\sum_{j=1}^ns_j)}}{k^2(k+1)(k+2)}\frac{\Pi_{j=1}^n{g_j \choose s_j}}
{{k \choose \sum_{j=1}^ns_j}}\nonumber \\
&\sum_{l=0}^{k}(k-l)\left\{\left[2k+(k-2)l\right]\Psi\left(k,\sum_{j=1}^ns_j,l\right)+\left[k^2-(k-2)l\right]\Phi\left(k,\sum_{j=1}^ns_j,l\right)\right\} \nonumber
\end{align}\end{linenomath}
where
\begin{linenomath}\begin{align}
&\Psi(k,i,l)={l \choose k-1-i}\frac{1}{(k-2)(k-1)^l}+{k-1-l \choose k-i}\frac{1}{(k-1)^{k-1-l}}, \nonumber\\
&\Phi(k,i,l)={l \choose k-i}\frac{1}{(k-1)^l}+{k-1-l \choose k-1-i}\frac{1}{(k-2)(k-1)^{k-1-l}}. \nonumber
\end{align}\end{linenomath}
$a_{s_1s_2\cdots s_n}-b_{(g_1-s_1)(g_2-s_2)\cdots (g_n-s_n)}$ in Eq~(\ref{sigma}) indicates the ``gains from flipping" \cite{2016-Pena-p1-15,2016-Pena-p-}, the change in payoffs for a focal A-player who interacts with $s_i$ A-players of type $i$ ($1\le i\le n$) in a group when all individuals change their strategies (from strategy A to strategy B or B to A) simultaneously.
Considering $\sum_{s_1=0}^{g_1}\sum_{s_2=0}^{g_2}\cdots\sum_{s_n=0}^{g_n}\sigma_{s_1s_2\cdots s_n}=1$, $\sigma_{s_1s_2\cdots s_n}$ can be viewed as a probability corresponding to term $a_{s_1s_2\cdots s_n}-b_{(g_1-s_1)(g_2-s_2)\cdots (g_n-s_n)}$.
Eq~(\ref{sigma}) thus indicates that strategy A is favored over B if the expected gain in payoffs from flipping is positive.
When $n=1$, our analytical prediction is fully line with a previous study about evolutionary multiplayer games on graphs \cite{2016-Pena-p1-15} (see S1 Text, Section 2).
To understand the structure coefficient for the case with $n>1$, we set the sum of the number of opposing A-players to be $S$, i.e., $\sum_{j=1}^ns_j=S$.
We find that $\sigma_{s_1s_2\cdots s_n}$ is the product of the structure coefficient corresponding to $n=1$ (denoted $\sigma_S$) and an additional term $\Pi_{j=1}^n{g_j \choose s_j}/{k \choose \sum_{j=1}^ns_j}$.
This term represents the probability of the configuration $s_1s_2\cdots s_n$ to occur under a given $S$.
Intuitively, with edge diversity, we distinguish A-players in the neighborhood by their types.
For a given number of A-players $S$, the probability of a specific configuration ($s_i$ A-players within $g_i$ individuals of type $i$) indeed follows the multivariate hypergeometric distribution $\Pi_{j=1}^n{g_j \choose s_j}/{k \choose S}$.
Our result shows that the structure coefficient associated with a specific configuration for diverse edges is simply a product of the probability for this configuration to occur and the corresponding structure coefficient without distinguishing edges.

Infinite populations usually serve as a baseline model to investigate the evolutionary dynamics of a system.
Therefore we conduct a consistent investigation in infinite populations.
The evolutionary dynamics of multiplayer games on graphs with edge diversity can be described in terms of replicator equation \cite{2006-Ohtsuki-p86-97} (see S1 Text, Section 4), given by
\begin{linenomath}\begin{align}
\dot{x}=\frac{\omega(k-2)x(1-x)}{k^2}f(x) \label{replicator_equation1},
\end{align}\end{linenomath}
where $f(x)$ is shown in Methods.
This seemingly complicated Eq~(\ref{replicator_equation1}) could be greatly simplified when applied to specific examples, such as traditional multiplayer games or pairwise games on graphs \cite{2006-Ohtsuki-p86-97}.
In the following, we apply Eq~(\ref{sigma}) and Eq~(\ref{replicator_equation1}) to several representative evolutionary scenarios, which cannot be dealt with by prior models.

\subsection{Applications}

When strategy A represents cooperation and B defection (A-players cooperators and B-players defectors),
Eq~(\ref{sigma}) can effectively predict the success of cooperation over defection in various interaction scenarios of multiplayer games such as volunteer's dilemmas \cite{1985-Diekmann-p605-610}, multiplayer stag-hunt game \cite{2009-Pacheco-p315-321}, and multiplayer snowdrift game \cite{2009-Souza-p581-588}.
Here we start with the prevailing collective activity in social insects and human societies---division of labor.

\textbf{Example 1. Evolutionary multiplayers games with division of labor.}
Consider a team of army ants retrieving prey items.
They can do this successfully only if different kinds of ants coordinate to perform corresponding subtasks \cite{2001-Franks-p635-642}.
In other words, cooperation from each kind of individuals is required to produce public goods.
We consider the simplest case with two kinds of individuals and the production of benefits requires at least one cooperator within each kind.
We use two types of edges on graphs to model this case: edges of type $1$ link the same kind of individuals and edges of type $2$ link different kinds of individuals.
A player obtains benefits only if in its neighborhood there are cooperative individuals along two types of edges.
Here we consider the evolution of individuals' behaviors (cooperation and defection) while remain individuals subtasks fixed throughout the evolution.
Payoff values are given by
\begin{equation} \label{MTPGG_na}
a_{s_1s_2} = \left\{
\begin{array}{ccl}
(s_1+s_2+1)\mathcal{B}-\mathcal{C} & & {s_2\ge1,}\\
-\mathcal{C} & & {\text{otherwise},}
\end{array} \right.
\end{equation}
\begin{equation} \label{MTPGG_nb}
b_{s_1s_2} = \left\{
\begin{array}{ccl}
(s_1+s_2)\mathcal{B} & & {s_1\ge1, s_2\ge1,}\\
0 & & {\text{otherwise},}
\end{array} \right.
\end{equation}
where $\mathcal{C}$ means the personal cost for each cooperator and $\mathcal{B}$ is the benefit to each participant.
Note that $a_{01}$ is not necessarily identical to $a_{10}$.
The public goods increase linearly with the number of cooperators, inasmuch as the number exceeds the corresponding threshold, termed accumulative effects of payoffs.
Substituting Eqs~(\ref{MTPGG_na}) and (\ref{MTPGG_nb}) into Eq~(\ref{sigma}), we have the critical benefit-to-cost ratio $(\mathcal{B}/\mathcal{C})^*$ above which cooperation is favored over defection, given by
\begin{linenomath}\begin{align}
(\mathcal{B}/\mathcal{C})^*=\frac{1}{\sum_{s_1=0}^{g_1}\sum_{s_2=1}^{g_2}(s_1+s_2+1)\sigma_{s_1s_2}-\sum_{s_1=0}^{g_1-1}\sum_{s_2=0}^{g_2-1}(k-s_1-s_2)\sigma_{s_1s_2}}. \nonumber
\end{align}\end{linenomath}
Then we consider the scenario without division of labor.
That is, benefits are produced as long as the total number of cooperators reaches a threshold.
For comparison, we set the threshold to be $2$.
Payoffs are thus $a_s=(s+1)\mathcal{B}-\mathcal{C}$ if $s\ge1$ and $a_s=-\mathcal{C}$ otherwise; $b_s=s\mathcal{B}$ if $s\ge2$ and $b_s=0$ otherwise.
$(\mathcal{B}/\mathcal{C})^*$ derived from Eq~(\ref{sigma}) is
\begin{linenomath}\begin{align}
(\mathcal{B}/\mathcal{C})^*=\frac{1}{\sum_{s=1}^{k}(s+1)\sigma_s-\sum_{s=0}^{k-2}(k-s)\sigma_s}.
\nonumber
\end{align}\end{linenomath}

\begin{figure*}[!h]
\centering
\includegraphics[width=1\textwidth]{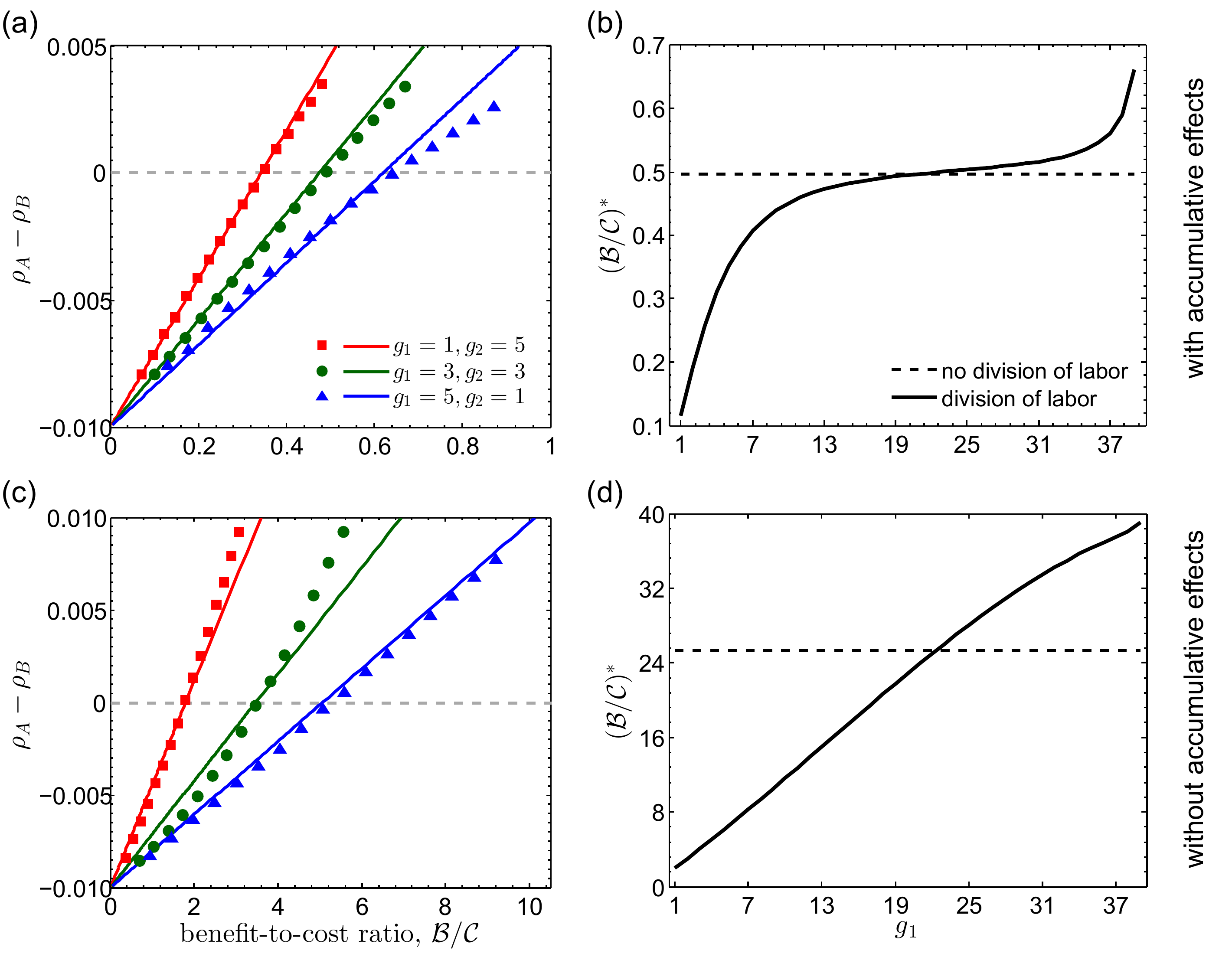}
\caption{\label{Fig 2} \textbf{Difference in fixation probability $\rho_\text{A}$-$\rho_\text{B}$ and critical benefit-to-cost ratio $(\mathcal{B}/\mathcal{C})^*$ for $\rho_{\text{A}}>\rho_{\text{B}}$ as a function of $g_1$}.
(ab) Division of labor with accumulative effects of payoffs (the increasing number of cooperators leads to the increasing productivity).
(cd) Division of labor without accumulative effects of payoffs (the productivity remains unchanged as the increasing number of cooperators).
In (a) and (c), we consider $n=2$ and different parameters of $g_1$ and $g_2$.
Dots presents simulation data (see Methods for simulation details) and lines are analytical predictions.
$\rho_{\text{A}}-\rho_{\text{B}}$ is analytically predicted by the product of the left side of Eq~(\ref{sigma}) and  the selection intensity $\omega$.
In (b) and (d), dash lines correspond to $(\mathcal{B}/\mathcal{C})^*$ for the case with no division of labor and solid lines $(\mathcal{B}/\mathcal{C})^*$ for the case with division of labor, where $g_2=40-g_1$.
}
\end{figure*}

Furthermore, we explore the case that the public goods remain fixed as the number of cooperators increases, inasmuch as the number exceeds the corresponding threshold (thus without accumulative effects of payoffs).
Payoffs are given by
\begin{equation}
a_{s_1s_2} = \left\{
\begin{array}{ccl}
\mathcal{B}-\mathcal{C} & & {s_2\ge1,} \nonumber\\
-\mathcal{C} & & {\text{otherwise},} \nonumber
\end{array} \right.
\end{equation}
\begin{equation} \label{MTPGG_b}
b_{s_1s_2} = \left\{
\begin{array}{ccl}
\mathcal{B} & & {s_1\ge1, s_2\ge1,} \nonumber\\
0 & & {\text{otherwise}.} \nonumber
\end{array} \right.
\end{equation}
We thus have
\begin{linenomath}\begin{align}
(\mathcal{B}/\mathcal{C})^*=\frac{1}{\sum_{s_1=0}^{g_1}\sum_{s_2=1}^{g_2}\sigma_{s_1s_2}-\sum_{s_1=0}^{g_1-1}\sum_{s_2=0}^{g_2-1}\sigma_{s_1s_2}}.
\nonumber
\end{align}\end{linenomath}
Analogously, in the counterpart with no labor division, if we set a single threshold 2, payoffs are $a_s=\mathcal{B}-\mathcal{C}$ if $s\ge1$ and $a_s=-\mathcal{C}$ otherwise; $b_s=\mathcal{B}$ if $s\ge2$ and $b_s=0$ otherwise.
We have $(\mathcal{B}/\mathcal{C})^*$
\begin{linenomath}\begin{align}
(\mathcal{B}/\mathcal{C})^*=\frac{1}{\sum_{s=1}^{k}\sigma_s-\sum_{s=0}^{k-2}\sigma_s}.
\nonumber
\end{align}\end{linenomath}
Panels Fig~\ref{Fig 2}a and \ref{Fig 2}c show that analytical predictions of fixation probabilities are in good agreement with results by Monte Carlo simulations for the whole range of benefit-to-cost ratios and for different parameters of $g_1$ and $g_2$.
In Fig~\ref{Fig 2}b, we show that with division of labor, $(\mathcal{B}/\mathcal{C})^*$ is a monotonous function of $g_1$.
Surprisingly, for small $g_1$, i.e., $g_1=1$, $(\mathcal{B}/\mathcal{C})^*$ is much lower than that without introducing division of labor.
Furthermore, for large $g_2$, i.e., $g_1=39$, $(\mathcal{B}/\mathcal{C})^*$ is far larger than that without introducing division of labor.
Therefore, the introduction of division of labor could significantly lower the barrier to establish a cooperative society, given a small number of individuals belong to the same type.
These findings are further confirmed when the increasing cooperation does not lead to the increasing productivity (see Fig~\ref{Fig 2}d).

\begin{figure*}[!h]
\centering
\includegraphics[width=0.9\textwidth]{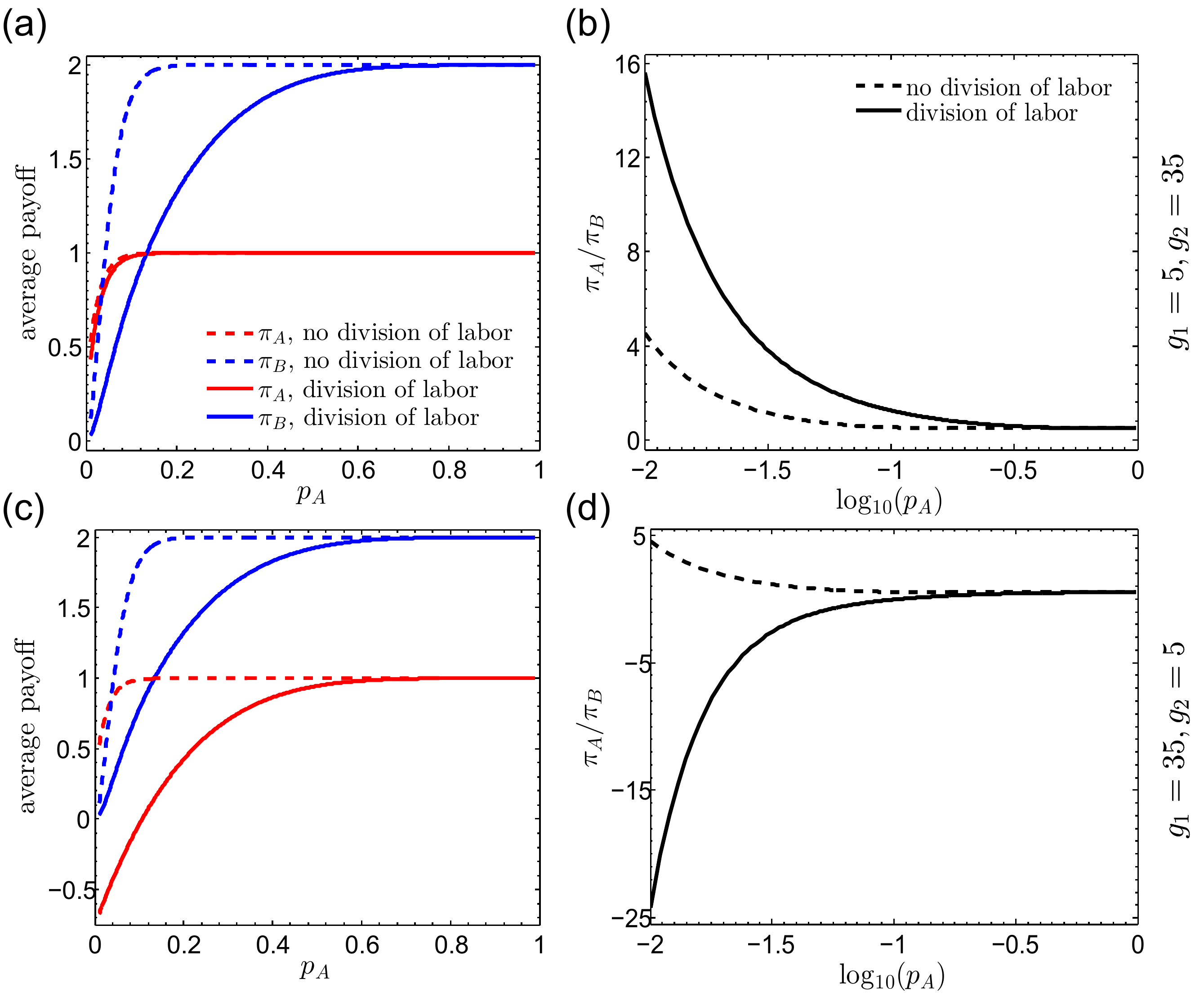}
\caption{\label{Fig 3} \textbf{Average payoffs as a function of $p_{\text{A}}$ with no (dash lines) and with (solid lines) division of labor.}
Here the increasing number of cooperators does not lead to increasing productivity, inasmuch as the number exceeds the threshold.
(a) For $g_1=5$ and $g_2=35$, division of labor does not affect the average payoff of A-players ($\pi_{\text{A}}$) much while reduce the average payoff of B-players ($\pi_{\text{B}}$) significantly.
This increases $\pi_{\text{A}}/\pi_{\text{B}}$ for the whole range of $p_{\text{A}}$ (b), and thus weakens the advantages of defectors over cooperators.
(c) For $g_1=35$ and $g_2=5$, division of labor reduces both $\pi_{\text{A}}$ and $\pi_{\text{B}}$ remarkably.
Nevertheless, the impact to $\pi_{\text{A}}$ is more noticeable than to $\pi_{\text{B}}$ (d) and thus reinforces the advantages of defectors over cooperators.
We take $\mathcal{B}=2$ and $\mathcal{C}=1$.
}
\end{figure*}

To make these explicit, we consider a case with a sufficiently large $k$ and without accumulative effects of payoffs.
With no division of labor (abbreviated to ``ndol''), payoffs of A- and B-players are respectively given by
\begin{linenomath}\begin{align}
\pi_{\text{A}}^{\rm ndol} &= \left[1-r^k(1-p_{\text{A}})^{k}\right]\mathcal{B}-\mathcal{C}, \nonumber\\
\pi_{\text{B}}^{\rm ndol} &= \left[1-(1-rp_{\text{A}})^{k-1}\left(1+kp_{\text{A}}-2p_{\text{A}}\right)\right]\mathcal{B}, \nonumber
\end{align}\end{linenomath}
where $r=(k-2)/(k-1)$ and $p_{\text{A}}$ is the fraction of A-players.
With division of labor (abbreviated to ``dol''), payoffs of A- and B-players are
\begin{linenomath}\begin{align}
\pi_{\text{A}}^{\rm dol} &= \left[1-r^{g_2}(1-p_{\text{A}})^{g_2}\right]\mathcal{B}-\mathcal{C}, \nonumber\\
\pi_{\text{B}}^{\rm dol} &= \left[1-(1-rp_{\text{A}})^{g_1}\right]\left[1-(1-rp_{\text{A}})^{g_2}\right]\mathcal{B}. \nonumber
\end{align}\end{linenomath}
For $0<p_{\text{A}}<1$, we have $\pi_{\text{A}}^{\rm dol}<\pi_{\text{A}}^{\rm ndol}$ and $\pi_{\text{B}}^{\rm dol}<\pi_{\text{B}}^{\rm ndol}$.
Thus division of labor transiently reduces the average payoffs of both A- and B-players.
This result is understandable since with the labor division the condition of producing benefits becomes more stringent.
However, in terms of the long-term development and stable states, the labor division is beneficial to the evolving system.
The labor division actually influence the competition between different behavioral traits and ultimately contributes to a cooperative society, which appears to be more prosperous.
To evaluate how the division of labor influences the competition between A- and B-players, we compare $\pi_{\text{A}}^{\rm dol}/\pi_{\text{B}}^{\rm dol}$ with $\pi_{\text{A}}^{\rm ndol}/\pi_{\text{B}}^{\rm ndol}$.
If $\pi_{\text{A}}^{\rm dol}/\pi_{\text{B}}^{\rm dol}> \pi_{\text{A}}^{\rm ndol}/\pi_{\text{B}}^{\rm ndol}$ ($\pi_{\text{A}}^{\rm dol}/\pi_{\text{B}}^{\rm dol}< \pi_{\text{A}}^{\rm ndol}/\pi_{\text{B}}^{\rm ndol}$), division of labor enhances (weakens) the advantage of A-players relative to B-players compared with that under no division of labor.
For $g_1\ll k$, $\pi_{\text{A}}^{\rm dol}$ approaches to $\pi_{\text{A}}^{\rm ndol}$, indicating the impact of division of labor to A-players is negligible (see Fig~\ref{Fig 3}a).
$\pi_{\text{B}}^{\rm dol}$ is the product of two terms (except $\mathcal{B}$).
One term, $1-(1-rp_{\text{A}})^{g_2}$, corresponds to the probability that there are cooperators among players belonging to a different type, roughly approximating to $\pi_{\text{B}}^{\rm ndol}/\mathcal{B}$.
The other term is the probability that there exist cooperators among players whose types are the same as the focal player.
For $g_1\ll k$, this term dominates the loss to B-players and weakens the advantages of defectors over cooperators.
The form of $\pi_{\text{B}}^{\rm dol}$ implies that division of labor essentially decomposes a many-player game into two fewer-player games, i.e., one game in which all participants show the same type as the focal player and one game in which participants' types are different from the focal player.
When the focal player belongs to a smaller group, it is harder to free-ride on others, which makes clear positive effects of division of labor on cooperation thriving.
Scenarios for $g_2\ll k$ can be analyzed analogously (see panels Fig~\ref{Fig 3}c and \ref{Fig 3}d).
This conclusion is still true with $n>2$ types of edges (see S2 Fig).
Our results suggest that the more specialized individuals are, namely, the less individuals are of the same type, the more cooperation will be achieved.
This may explain the flourishing cooperation in the highly specialized human societies.

\textbf{Example 2. Diverse multiplayer games.}
We investigate a scenario where individuals are engaged in different games concurrently, irrespective of two-player or multiplayer games.
We let individuals linked by the same type of edges form a group to play a multiplayer game.
This means that each focal individual participates in $n$ multiplayer games.
These games can differ in payoff structures, i.e., game metaphors and payoff values.
We assume that any two games are independent and each player accumulates its payoffs gained from each game, i.e.,
\begin{linenomath}\begin{align}
&a_{s_1s_2\cdots s_n}=a_{s_1}^1+a_{s_2}^2+\cdots+a_{s_n}^n, \nonumber \\ %\label{application_a} \\
&b_{s_1s_2\cdots s_n}=b_{s_1}^1+b_{s_2}^2+\cdots+b_{s_n}^n. \nonumber  %\label{application_b}
\end{align}\end{linenomath}
where $a_{s_i}^i$ ($b_{s_i}^i$) presents the payoff assigned to an A-player (a B-player) in the interaction with individuals of type $i$ when there are $s_i$ opposing A-players.
If $g_1=g_2=\cdots=g_n=g$, Eq~(\ref{sigma}) can be simplified as (see S1 Text, Section 3)
\begin{linenomath}\begin{align} \label{sigma_g1g2}
\sum_{s=0}^g\tilde{\sigma}_s\left(\sum_{j=1}^n a_{s}^j-\sum_{j=1}^n b_{g-s}^j\right)>0
\end{align}\end{linenomath}
where $\tilde{\sigma}_s=\sum_{s_2=0}^{g_2}\sum_{s_3=0}^{g_3}\cdots\sum_{s_n=0}^{g_n}\sigma_{ss_2\cdots s_n}$.
For such a system, we just need $g+1$ structure coefficients to describe the effects of population structures on the evolution of two traits.
If designating
\begin{linenomath}\begin{align}
&\bar{a}_{s}=\frac{1}{n}\sum_{j=1}^n a_{s}^j, \nonumber\\
&\bar{b}_{s}=\frac{1}{n}\sum_{j=1}^n b_{s}^j, \nonumber
\end{align}\end{linenomath}
We have the condition for $\rho_A>\rho_B$, given by
\begin{linenomath}\begin{align} \label{sigma_average}
\sum_{s=0}^g\tilde{\sigma}_s\left(\bar{a}_{s}-\bar{b}_{g-s}\right)>0.
\end{align}\end{linenomath}
Note that $\bar{a}_{s}$ ($\bar{b}_{s}$) corresponds to the payoff averaged over all games when there are $s$ opposing cooperators.
This suggests that while payoff structures are diverse in different interactions, the evolutionary outcome can be predicted by assuming that all interactions are governed by a unified payoff structure, i.e., the `average' over all structures.
Alternatively, we can rewrite Eq~(\ref{sigma_g1g2}) as $\sum_{j=1}^n\left[\sum_{s=0}^g\tilde{\sigma}_s\left(a_{s}^j-b_{g-s}^j\right)\right]>0$.
Note that $\sum_{s=0}^g\tilde{\sigma}_s\left(a_{s}^j-b_{g-s}^j\right)$ presents the results when all interactions described by the single payoff structure, i.e., $a_{s}^j$ and $b_{s}^j$.
Therefore, the evolutionary outcome under diverse multiplayer games can be viewed as the sum of results obtained when all interactions are governed by a single payoff structure.
Both the two interpretations significantly simplify the calculation complexity when the payoff forms are relation-dependent.
We further confirm the above findings in infinite populations (see S1 Text, Section 3).

\begin{figure*}[!h]
\centering
\includegraphics[width=0.71\textwidth]{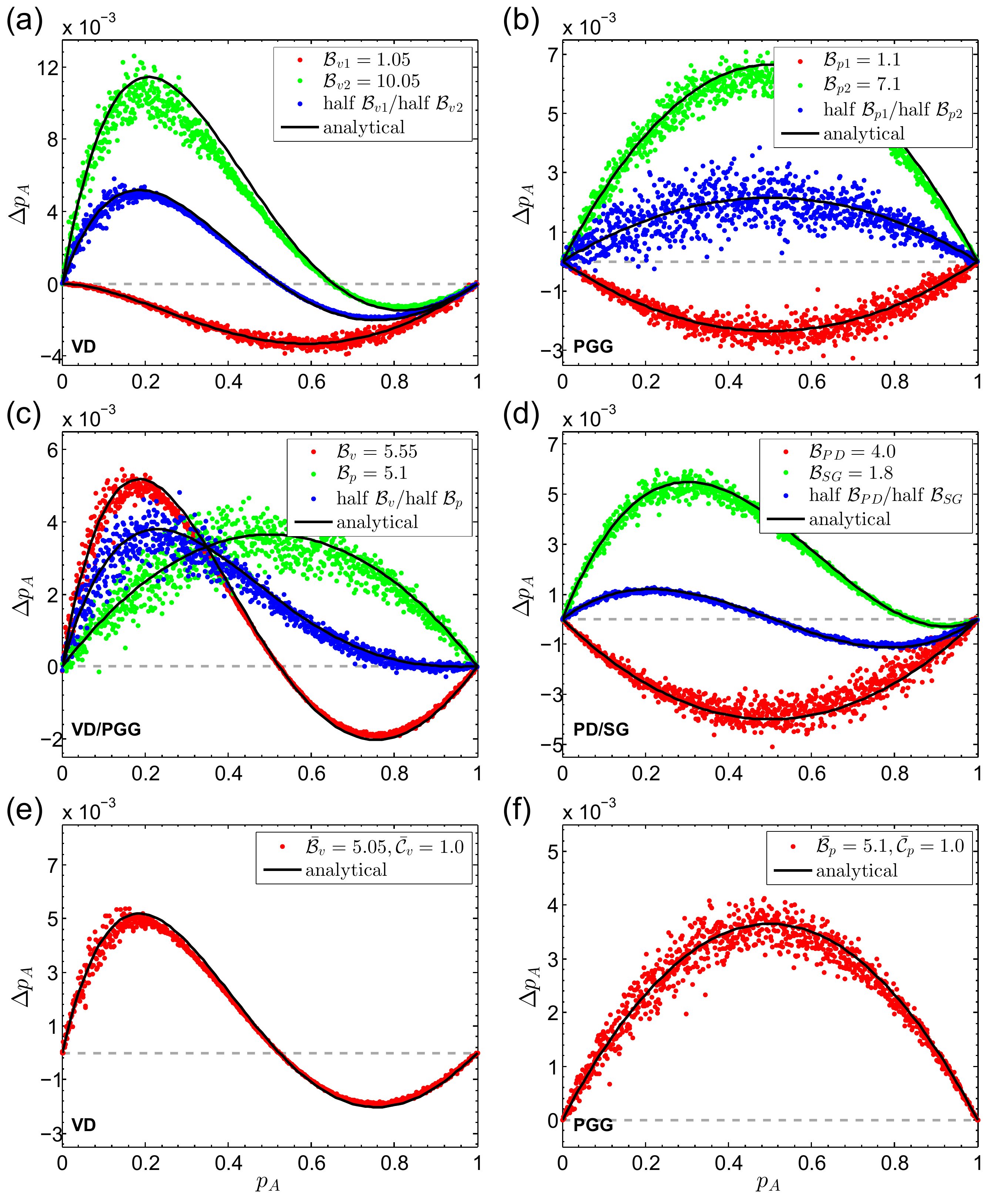}
\caption{\label{Fig 4} \textbf{Average change ($\Delta p_{\text{A}}$) in the frequency of A-players ($p_{\text{A}}$), in volunteer's dilemma (\textbf{VD}), public goods games (\textbf{PGG}), diverse multiplayer games (\textbf{VD/PGG}), and diverse two-player games (\textbf{PD/SG})}.
The population structure is a random regular graph with $N=1000$, $n=2$, and $g_1=g_2=3$.
In (a-c), the cost to cooperate is fixed to $1$ and benefits are shown in the legend of each panel.
In (d), under \textbf{PD}, a cooperator bears a cost $1$ to provide its opponent with a benefit $\mathcal{B}_{PD}$.
Under each \textbf{SG}, the total cost for cooperators is $1$ and the benefit for each player is $\mathcal{B}_{SG}$.
Three cases are investigated in each panel of (a-d).
For example, in (a), benefits in all multiplayer interactions are $\mathcal{B}_{v1}=1.05$ (red dots), $\mathcal{B}_{v2}=10.05$ (green dots), or designated at equal proportions (blue dots).
In (ef), both benefits and costs in each interaction are sampled according to a Gaussian distribution, with mean $5.05$, variance $1.5$ for benefits and mean $1.0$, variance $0.25$ for costs (e), mean $5.1$ and variance $1.5$ for benefits and mean $1.0$ and variance $0.25$ for costs (f).
Dots represent the simulation data and lines are analytical predictions based on unified interactions with average payoffs (see Methods for simulation details).
}
\end{figure*}

We illustrate a few examples in Fig~\ref{Fig 4}, including nonlinear multiplayer game like volunteer dilemmas \cite{1985-Diekmann-p605-610} and linear public goods games.
In a volunteer dilemmas, once an individual volunteers by bearing a cost $\mathcal{C}_v$, each participant obtains a benefit $\mathcal{B}_v$.
In Fig~\ref{Fig 4}a, each individual participates in two volunteer dilemmas in each generation.
When $\mathcal{B}_v=1.05$ and $\mathcal{C}_v=1$ in one interaction and $\mathcal{B}_v=10.05$ and $\mathcal{C}_v=1$ in the other, the evolutionary dynamics can be approximated by the case where all interactions are described by a unified game with $\mathcal{B}_v=(1.05+10.05)/2$ and $\mathcal{C}_v=(1+1)/2$.
Alternatively, dynamics for the case with half $\mathcal{B}_{v1}=1.05$ and half $\mathcal{B}_{v2}=10.05$ (blue) can be viewed as the average over that with full $\mathcal{B}_{v1}=1.05$ (red) and that with full $\mathcal{B}_{v2}=10.05$ (green).
Panels Fig~{\ref{Fig 4}}b and {\ref{Fig 4}}c confirm above findings in linear public goods games and mixed games (half volunteer dilemmas and half linear public goods games).

We highlight above rules can be further extended to more general cases.
When each collective interaction is endowed with an independent payoff structure (payoff structures in any two interactions centered on player $x$ are independent; besides, payoff structures in any interaction centered on player $x$ and those centered on $y$ are uncorrelated),
the collective behavior still can be predicted by an `average' case over all interactions (see panels Fig~\ref{Fig 4}e and \ref{Fig 4}f).
Furthermore, if the numbers of participants in different collective interactions are not identical, interactions with the same number of participants can be described by their `average' case.
That is, if $g_{l_1}=g_{l_2}=\cdots=g_{l_u}\ne g_{m_1}=g_{m_2}=\cdots=g_{m_v}$, interactions with individuals belonging to type $l_1$, $l_2$, $\cdots$, $l_u$ can be resolved as uniform interactions with payoff matrix $\bar{a}_{s}^l=\sum_{j=l_1}^{l_u} a_{s}^j/u$ and $\bar{b}_{s}^l=\sum_{j=l_1}^{l_u} b_{s}^j/u$.
Interactions associated with edges of type $m_1$, $m_2$, $\cdots$, $m_v$ can be treated as uniform interactions with payoff matrix $\bar{a}_{s}^m=\sum_{j=m_1}^{m_v} a_{s}^j/v$ and $\bar{b}_{s}^m=\sum_{j=m_1}^{m_v} b_{s}^j/v$, applicable to sufficiently large finite and infinite populations.
Generally, if there are $m$ different game sizes among $n$ multiplayer games, i.e., $g_1$, $g_2$, $\cdots$, $g_m$, satisfying $g_i\neq g_j$ if $i\neq j$ ($1\le i,j\le m$), the number of structure coefficients needed to describe the effects of population structures decreases to
$\sum_{i=1}^m(g_i+1)$.
Therefore, in the absence of edge diversity (thus $m=1$ and $g_1=k$), the number of structure coefficients is $k+1$, in line with a previous study \cite{2016-Pena-p1-15}.
%If all interactions goes on between two players (thus $m=1$ and $g_1=1$), the number of structure coefficients is $2$.
If game sizes for all multiplayer games are different (thus $m=n$), we need $\sum_{i=1}^n(g_i+1)$ to predict the evolutionary outcome.
Table~\ref{Table 3} summarizes the number of structure coefficients in various cases.

\begin{table}
%\begin{adjustwidth}{-2.25in}{0in} % Comment out/remove adjustwidth environment if table fits in text column.
\caption{\label{Table 3}
\textbf{The number of structure coefficients to predict the evolutionary outcome.} } \centering
\begin{tabular}{p{4cm}<{\centering}| p{4cm}<{\centering} p{8cm}<{\centering} }
\hline
\hline
            & general payoff structure & payoff structure of diverse multiplayer games \\
\hline
general spatial structure & $\Pi_{i=1}^m{g_i+n_i \choose n_i}$ & $\sum_{i=1}^m(g_i+1)$ \\
\hline
$g_i=g$ for any $1\le i\le n$ & ${g+n \choose n}$ & $g+1$ \\
\hline
$g_i\ne g_j$ for any $i\ne j$ & $\Pi_{i=1}^n(g_i+1)$ & $\sum_{i=1}^n(g_i+1)$\\
\hline
\hline
\end{tabular}
\begin{flushleft} In the general spatial structure, there are $m$ different values among all $g_i$s ($1\le i\le n$).
We denote $g_1$, $g_2$, $\cdots$, $g_m$ these values and $n_i$ the number of value $g_i$, i.e., $k=\sum_{i=1}^m n_ig_i$.
Note that we can further eliminate an extra structure coefficient through dividing the sigma rule [see Eq~(\ref{sigma})] by a positive structure coefficient.
\end{flushleft}
%\end{adjustwidth}
\end{table}

\textbf{Example 3. Evolutionary multiplayer games on weighted graphs.}
We proceed with the application of above findings on weighted graphs.
Interactions between individuals often differ in capacity, frequency, and strength \cite{2004-Barrat-p3747-3752}.
Weighted graphs well incorporate these factors where weights of edges are proportional to interaction frequencies.
Partly since the simple and intuitive understanding of weighted edges, most studies about games on weighted graphs so far are based on two-player interactions \cite{2007-Taylor-p469-469,2014-Debarre-p3409-3409,2014-Allen-p113-151,2017-Allen-p227-227,2018-Zhou-p-}.
Although collective interactions can also occur at different interaction rates like two-player versions, few studies explore it.
The framework proposed in this paper is also applicable to investigate the multiplayer games on weighted graphs, where different group interactions occur at different rates.
Concretely, individuals linked by the same type of edges are engaged in a group interaction and these edges are endowed with a uniform weight which represents the frequency of this group interaction.
Thus, a larger value of edge weight means the more frequent contact \cite{2007-Taylor-p469-469,2014-Debarre-p3409-3409,2014-Allen-p113-151,2017-Allen-p227-227,2018-Zhou-p-} or more diffusible public goods between interactants \cite{2013-Allen-p1169-1169,2018-Su-p-}.
Counter-intuitively, we show that strong social ties do not change the evolutionary fate of cooperation, irrespective of based on multiplayer or two-player games (see S1 Text, Section 4).
As shown in Fig~\ref{Fig 5}a, in finite populations, strong social ties just amplify the fixation probability (both $\rho_A$ and $\rho_B$) while remains the critical condition $\mathcal{B}/\mathcal{C}$ for $\rho_A>\rho_B$ unchanged.
Analogously, in infinite populations, strong social ties accelerate the evolutionary rate which do not change the inner equilibria at all (Fig~\ref{Fig 5}b).
We can make this clear by virtue of conclusions in Example 2.
In volunteer's dilemmas, the payoff matrix for interactions with individuals of type $j$ is $a_{s}^j=\mathcal{B}_v^j-\mathcal{C}_v^j$ for any $s$, $b_{s}^j=\mathcal{B}_v^j$ for $s>0$, and $b_{s}^j=0$ for $s=0$.
We take $\mathcal{B}_v^j=\zeta_j\mathcal{B}_v$ and $\mathcal{C}_v^j=\zeta_j\mathcal{C}_v$, where $\zeta_j$ denotes the weight of edges linking individuals of type $j$.
From Example 2, the evolutionary dynamics can be approximated by unified interactions with payoff matrix $\bar{a}_{s}=\bar{\mathcal{B}}_v-\bar{\mathcal{C}}_v$ for any $s$, $\bar{b}_{s}=\bar{\mathcal{B}}_v$ for $s>0$, and $\bar{b}_{s}=0$ for $s=0$, where $\bar{\mathcal{B}}_v=\sum_{j=1}^n\mathcal{B}_v^j/n=\mathcal{B}_v\sum_{j=1}^n\zeta^j/n$ and $\bar{\mathcal{C}}_v=\mathcal{C}_v\sum_{j=1}^n\zeta^j/n$.
Combining Eq~(\ref{sigma_average}), we have the critical condition
\begin{linenomath}\begin{align}
\left(\frac{\mathcal{B}_v}{\mathcal{C}_v}\right)^*=\frac{1}{\tilde{\sigma}_g}\nonumber
\end{align}\end{linenomath}
above which $\rho_A>\rho_B$.
Note that $\left(\mathcal{B}_v/\mathcal{C}_v\right)^*$ is independent of edge weights $\sum_{j=1}^n\zeta^j$.

\textbf{Example 4. Evolutionary two-player games on graphs with edge diversity.}
As a consistency check, we investigate two-player games.
Distinguished from previous studies, %here each individual can play different two-player games with different neighbors.
here each type of edges are endowed with an independent payoff matrix.
The payoff matrix for interactions occurring in edges of type $i$ is
\begin{linenomath}
$$\bordermatrix{
  & \text{A} & \text{B} \cr
\text{A} & \alpha_i & \beta_i \cr
\text{B} & \gamma_i & \theta_i \cr
},$$
\end{linenomath}
where each value corresponds to the payoff assigned to the individual adopting a strategy in the row against its partner taking a strategy in the column.
Transforming the payoff to multiplayer interactions through
$a_{s_1s_2\cdots s_n}=\sum_{i=1}^n \left[s_i\alpha_i+(g_i-s_i)\beta_i\right]$ and
$b_{s_1s_2\cdots s_n}=\sum_{i=1}^n \left[s_i\gamma_i+(g_i-s_i)\theta_i\right]$, we have the sigma rule from Eq~(\ref{sigma})
\begin{linenomath}\begin{align}
\sum_{i=1}^n \bar{s}_i\alpha_i+\sum_{i=1}^n \left(g_i-\bar{s}_i\right)\beta_i
-\sum_{i=1}^n \left(g_i-\bar{s}_i\right)\gamma_i-\sum_{i=1}^n \bar{s}_i\theta_i>0 \nonumber,
\end{align}\end{linenomath}
where
\begin{linenomath}\begin{align}
\bar{s}_i = \sum_{s_1=0}^{g_1}\sum_{s_2=0}^{g_2}\cdots\sum_{s_n=0}^{g_n}\sigma_{s_1s_2\cdots s_n}s_i \nonumber.
\end{align}\end{linenomath}
Applying (see S1 Text, Section 3)
\begin{linenomath}\begin{align}
\sum_{s_1=0}^{g_1}\sum_{s_2=0}^{g_2}\cdots\sum_{s_n=0}^{g_n}\sigma_{s_1s_2\cdots s_n}s_i = \frac{g_i(k+1)}{2k}, \nonumber
\end{align}\end{linenomath}
we have the sigma rule for evolutionary two-player games on graphs with edge diversity
\begin{linenomath}\begin{align}
\sum_{i=1}^n\left[g_i(k+1)\alpha_i+g_i(k-1)\beta_i\right]>\sum_{i=1}^n\left[g_i(k-1)\gamma_i+g_i(k+1)\theta_i\right]. \nonumber
\end{align}\end{linenomath}
Since $\sum_{i=1}^n g_i=k$, dividing both sides of the above condition by $k$, we obtain the simplified condition
\begin{linenomath}\begin{align}
\frac{k+1}{k-1}\bar{\alpha}+\bar{\beta}>\bar{\gamma}+\frac{k+1}{k-1}\bar{\theta} \nonumber
\end{align}\end{linenomath}
where $\bar{\alpha}=(1/k)\sum_{i=1}^n g_i \alpha_i$, $\bar{\beta}=(1/k)\sum_{i=1}^n g_i \beta_i$, $\bar{\gamma}=(1/k)\sum_{i=1}^n g_i \gamma_i$, and $\bar{\theta}=(1/k)\sum_{i=1}^n g_i \theta_i$.
The above condition suggests that for pairwise games contingent on the edges, it suffices to study a unified game with its payoff entries averaged over all the games.
Note that for the unified game, the associated structure coefficient is $(k+1)/(k-1)$, which coincides with
that with $n=1$ \cite{2006-Ohtsuki-p502-502,2009-Tarnita-p570-581}.
Moreover, if all the games are in the form of donations games, i.e.,  $\alpha_i=\mathcal{B}_i-\mathcal{C}_i$, $\beta_i=-\mathcal{C}_i$, $\gamma_i=\mathcal{B}_i$, and $\theta_i=0$, the condition for natural selection favoring cooperation over defection is
\begin{linenomath}\begin{align} \label{BCk}
\frac{\bar{\mathcal{B}}}{\bar{\mathcal{C}}}>k.
\end{align}\end{linenomath}
where $\bar{\mathcal{B}}=(1/k)\sum_{i=1}^n g_i \mathcal{B}_i$ and $\bar{\mathcal{C}}=(1/k)\sum_{i=1}^n g_i \mathcal{C}_i$
This equation thus extends a well-known $\mathcal{B}/\mathcal{C}>k$ rule ($\mathcal{B}$ and $\mathcal{C}$ are respectively the benefit and cost of the donative behavior) \cite{2006-Ohtsuki-p502-502} to a general $\bar{\mathcal{B}}/\bar{\mathcal{C}}>k$ rule where $\bar{\mathcal{C}}$ means the average cost for cooperative behavior on all possible types of edges and $\bar{\mathcal{B}}$ is the average benefit \cite{2015-McAvoy-p1004349-1004349} (see Fig~\ref{Fig 4}f).

\section{Discussion}
Due to variations in both environment or gene, individuals own distinct social status or play different roles in colonies \cite{1964-Hamilton-p1-16,2013-Kun-p2453-2453}.
Typically, individuals with geographic proximity and genetic similarity tend to establish stronger social ties than those separated by remote geographic space or distinguished by large genetic difference.
Encountering different types of individuals, one may be affected differently.
Here we model the heterogeneous influence by different types of edges and develop a framework of evolutionary multiplayer games on graphs with edge diversity.
Since the two-player game is the simplest multiplayer game, our findings are applicable to pairwise interactions.
We make a thorough investigation in both finite and infinite populations.
We provide the analytical formulas of structure coefficients for random regular graphs with $n$ types of edges, which effectively predicts when natural selection favors one strategic behavior over the other.

As the first application of our framework, we consider how the division of labor affects the evolution of cooperation.
As well known, the division of labor prevails in colonies of social insects, hunting groups of lions, and human societies \cite{2014-Wright-p9533-9537,2001-Franks-p635-642,2005-Kay-p165-174,1986-Franks-p425-429,1992-Stander-p445-454},
where individuals are born or trained to perform specialized subtasks.
Such specialization not only makes them more productive on their own subtasks but also results in synergistic effects on the overall productivity when they cooperate with each other.
We here model the strategic interactions under the division of labor as a multi-threshold public goods game.
The public goods are provided only when individuals of distinct types cooperate.
We find that the division of labor could promote the evolution of cooperation.
The reason lies in that task specialization decomposes a many-player interaction into several fewer-player interactions.
Such a decomposition helps reduce the free-riding behaviors.
%[More explanations are needed].

Our framework are also able to address the more realistic situation where individuals concurrently face diverse social dilemmas.
This is in stark contrast with the ideal assumption in most previous studies where all interactions are described by a unified game metaphor \cite{2014-Li-p5536-5536,2015-Zhou-p60006-60006,2016-Li-p22407-22407,2016-Pena-p1-15,2015-Pena-p122-136,2013-Wu-p-,2015-Du-p8014-8014}.
In the real word, an individual may be caught in a volunteer's dilemma with its colleagues and meanwhile engage in public goods games with its neighbors.
The inevitable extinction of cooperation in the public goods game seems desperate.
Fortunately, the public goods game is merely one of the many types of social dilemmas individuals encounter.
Our work reveals that leveraging the distinct nature of diverse social dilemmas can entail an evolutionary outcome where cooperators are rescued and are able to coexist with defectors.
In addition, a seminal work by McAvoy \emph{et. al.} tells that under asymmetric two-player games the evolutionary processes behave macroscopically like that governed by symmetric games \cite{2015-McAvoy-p1004349-1004349}.
Here we confirm that irrespective of two-player or multiplayer games, the evolutionary dynamics with diverse interactions can be approximated by that governed by a single game.
For more complicated cases where sizes of group
interactions are different, we also provide an efficient method to simplify it.
Our work greatly reduces the complexity when investigating the evolutionary dynamics in real-world systems.

Besides, multiplayer games on weighted graphs can be considered.
We find that the presence of strong social ties does not always provide an evolutionary advantage to cooperators, which seems to coincide with recent findings under aspiration dynamics \cite{2018-Zhou-p-}.
This contrasts with the conclusion in Ref. \cite{2017-Allen-p227-227} where they show that strong ties boost cooperation most.
The main difference between our work and theirs is that we do not couple the strength of interactions and the probability of replacement along an edge.
In their work, a strong social tie indicates not only a higher frequency of interactions but also a more probable path for strategy dispersal.
Simultaneously enhancing the strength of interactions and the likelihood of dispersal lead to a strong strategy reciprocity between individuals and thus facilitate the clustering of cooperators.
However, if strong ties merely indicate frequent interactions as in our work, we show that they fail to promote cooperation, irrespective of group or pairwise interactions.
Note that in our model, individuals derive payoffs only from interactions with their nearest neighbors \cite{2002-Szabo-p118101-118101,2016-Pena-p1-15}.
When individuals can interact with both the nearest and second-nearest neighbors, the impact of social ties on the evolution of cooperation are more complicated \cite{2018-Su-p-}.
A further investigation along this direction may generate new insights.

Our work also extends the research scope about the interplay between the evolution of a population and the diversity.
The two basic elements of a population are individuals and social ties.
Most prior studies about diversity focus on individuals' attributes, such as the number of social ties they have, the ability to influence their opponents, etc \cite{2008-Santos-p213-213,2012-Santos-p88-96}.
Such diversity highlights that two individuals are different when possessing different attributes.
Here we stress the diversity of social ties.
Social ties not just establish the connections between separated individuals.
They carry a massive amount of information about two connected individuals, such as the intimacy of the interpersonal relationships, the frequency of physical contact, and even the history about previous interactions.
All these are unlikely to be captured by individuals' attributes.
The example of division of labor also proved that the diversity of social ties (or edge diversity) could catalyze cooperation.
Our recent work about interactive diversity is pertinent to this topic \cite{2017-Su-p103023-103023,2018-Su-p149-157}.
Interactive diversity describes that each individual adopts independent strategies in different interactions.
Thus even facing an identical strategy by two different opponents, the focal individual could be influenced differently due to its own behavior.
Nevertheless, the influence difference fully depends on strategies between interactants and is unrelated to other information like genetic similarity or geographic proximity.
Thus, interactive diversity does not essentially capture diverse social ties explored in this paper \cite{2016-Su-p103007-103007}.
We wish our work could attract more work into the evolutionary dynamics along edges.

In this paper we constrain that each social tie has symmetric effects on connected individuals.
For example, if Alice is close to Bob in consanguinity or geographic sites, Bob is close to Alice.
Thus the benefit that cooperative Alice brings to Bob is identical to that of cooperative Bob to Alice.
A promising and challenging extension is the interactions with asymmetric social ties, such as the relationship between leaders and followers.
In such case, each individual should be endowed with an independent payoff function \cite{2011-Gokhale-p180-191,2016-McAvoy-p203-238}.
Despite much complicity in analytical calculations, we expect a further research into this realistic situation, which is bound to provide fruitful insights.
We point out that our theoretical results are based on assumption of weak selection, as used by most previous theoretical studies \cite{2014-Li-p5536-5536,2015-Zhou-p60006-60006,2016-Li-p22407-22407,2016-Pena-p1-15,2016-Pena-p-,2015-Pena-p122-136,2015-Du-p8014-8014}.
Although the assumption of weak selection is reasonable in many cases and also make this conundrum accessible to analytical calculation \cite{2010-Wu-p46106-46106}, other situations routinely encountered in social or natural science are better captured by strong selection.
Thus, a further investigation with strong selection is necessary to enrich our understanding to the collective behavior in complex systems \cite{2011-Veelen-p116-128,2007-Traulsen-p522-529}.
Finally, in this paper, we assume that the types of edges remain unchanged throughout the evolution.
This is natural in many cases, like when types of edges indicate the geographic proximity.
Nevertheless, when edges' types represent the genetic difference between linked individuals and the population evolve based on individuals' reproduction, edges' types evolve as well \cite{2015-McAvoy-p1004349-1004349}.
A study into the coevolution of individuals' traits and edge types is expected.

\section*{Methods}
\subsection*{Theoretical analysis}
We derive the analytical formulas based on the combination of pair approximation and diffusion theory.
The method of pair approximation is formulated for infinite Cayley trees or Bethe lattices, which are regular graphs without any loops.
For finite but sufficiently large random regular graphs ($N \gg k$), loops tend to be quite large, which has negligible impacts to validity of the pair approximation.
Thus the obtained formulas approximate the simulated results.
The detailed theoretical derivations is provided in Supporting Information.
$f(x)$ in Eq~(\ref{replicator_equation1}) is
\begin{linenomath}\begin{align}
f(x)=
&\sum_{s_1=0}^{g_1}\sum_{s_2=0}^{g_2}\cdots \sum_{s_n=0}^{g_n}
\left[\prod_{j=1}^n {g_j \choose s_j}x^{s_j}(1-x)^{g_j-s_j}\right]\left(\Lambda_a-\Lambda_b\right), \nonumber
%\label{replicator_equation}
\end{align}\end{linenomath}
\begin{linenomath}\begin{align}
\Lambda_a=
&\sum_{r_1=0}^{g_1-s_1}\sum_{r_2=0}^{g_2-s_2}\cdots \sum_{r_n=0}^{g_n-s_n}
\left[\prod_{j=1}^n {g_j-s_j \choose r_j}z^{r_j}(1-z)^{g_j-s_j-r_j}\right]\nonumber \\
&\sum_{j=1}^n\Big[\left(s_j+r_j\right)a_{(s_1+r_1)(s_2+r_2)\cdots(s_n+r_n)}\nonumber \\
&\quad \quad +\left(zs_j+\frac{r_j}{z}\right)a_{(s_1+r_1-\delta_{1j})(s_2+r_2-\delta_{2j})\cdots(s_n+r_n-\delta_{nj})}\Big], \nonumber\\
\Lambda_b=
&\sum_{r_1=0}^{s_1}\sum_{r_2=0}^{s_2}\cdots \sum_{r_n=0}^{s_n}
\left[\prod_{j=1}^n {s_j \choose r_j}z^{r_j}(1-z)^{s_j-r_j}\right] \nonumber \\
&\sum_{j=1}^n\Big[\left(g_j-s_j+r_j\right)b_{(s_1-r_1)(s_2-r_2)\cdots(s_n-r_n)} \nonumber \\
&\quad \quad +\left(z(g_j-s_j)+\frac{r_j}{z}\right)b_{(s_1-r_1+\delta_{1j})(s_2-r_2+\delta_{2j})\cdots(s_n-r_n+\delta_{nj})}\Big]. \nonumber
\end{align}\end{linenomath}
$z=1/(k-1)$.
$\delta_{ij}$ equals to $1$ if $j=i$ and $0$ otherwise.

\subsection*{Computer simulations}
Network generation: We present the procedure to produce a random regular graph with $n$ types of edges, where the number of edges of type $i$ linked to each node is $g_i$ ($1\le i\le n$).
We take $g_1\ge 2$.
Given values of $g_i$, we first construct a random regular graph of degree $g_1$ and make sure that it is connected.
All edges in this graph are assigned to be type $1$.
Then we augment this graph by increasing the degree of all nodes by $g_2$.
All edges added in this step are assigned to be type $2$.
Repeating this procedure for $n-1$ times where the augment degree is $g_{i+1}$ in $i_{th}$ augment, we assign the edges added in $i_{th}$ augment to be type $i+1$.
Finally, we generate a random regular graph with degree $\sum_{i=1}^n g_i$.  \\
Fixation probability $\rho_{\text{A}}$: In a generated random regular graph with $N=200$ and $n=2$ ($g_1$ and $g_2$ are given in corresponding figures), a random node is selected to be A-player and the rest are B-players.
The system evolves as described in Models with selection intensity $\omega=0.01$.
The evolution does not end until all nodes turn to A-players or B-players.
Repeating graph generation and subsequent system evolution for $10^7$ runs, $\rho_{\text{A}}$ is the fraction of times where A-players reach fixations.
$\rho_{\text{B}}$ is calculated analogously.\\
%Results for $N=500$ are provided in the Supplementary Information.
Replicator equation: In a generated random regular graph with $N=1000$ and $g_1=3$, $g_2=3$, a random value of $f$ is sampled uniformly from the interval $[0,1]$.
Then each node is initiated to be a cooperator with probability $f$ and a defector otherwise.
The system evolves as described in Model with selection intensity $\omega=0.01$.
We term a time step during which the population updates $N$ times.
Let $p_{\text{A}}(t)$ denote the frequency of A-players at time step $t$ and $p_{\text{A}}(0)$ the initial frequency of A-players.
Let $\Delta p_{\text{A}}(t)$ denote the change in frequency of A-players within a time step starting at time step $t$, i.e.,
$\Delta p_{\text{A}}(t)=p_{\text{A}}(t+1)-p_{\text{A}}(t)$.
$\Delta p_{\text{A}}(t)$ is associated with $p_{\text{A}}(t)$ and is recorded.
The evolution does not end until all nodes turn to A-players or B-players.
The graph generation, sample of $f$, and subsequent system evolution, are repeated for $50000$ times if there is an inner equilibria, which can be predicted by Eq~(\ref{replicator_equation1}), and for $1000000$ times if there is no any inner equilibria.
Finally, $\Delta p_{\text{A}}$ corresponding to $p_{\text{A}}$ is the average of recorded $\Delta p_{\text{A}}(t)$, as plotted in Figs~\ref{Fig 4} and \ref{Fig 5}.

\section*{Acknowledgments}
We appreciate Alex McAvoy for insightful discussions.
This work is supported by the National Natural
Science Foundation of China (NSFC) under grant no. 61751301 and no. 61533001.
Q. S. acknowledges the support from China Scholarship Council (CSC) under no. 201706010277.
%The funders had no role in study design, data collection and analysis, decision to publish, or preparation of the manuscript.

%\bibliographystyle{apsrev4-1} % 控制文献显示格式的
%\bibliography{D:/RESEARCH_PROJECT/References/SQ12/PlOS_Computational_Biology1}  %保证 LaTeX 可以找到该 bib 文件

\clearpage

\begin{figure*}[!h]
\centering
\includegraphics[width=1\textwidth]{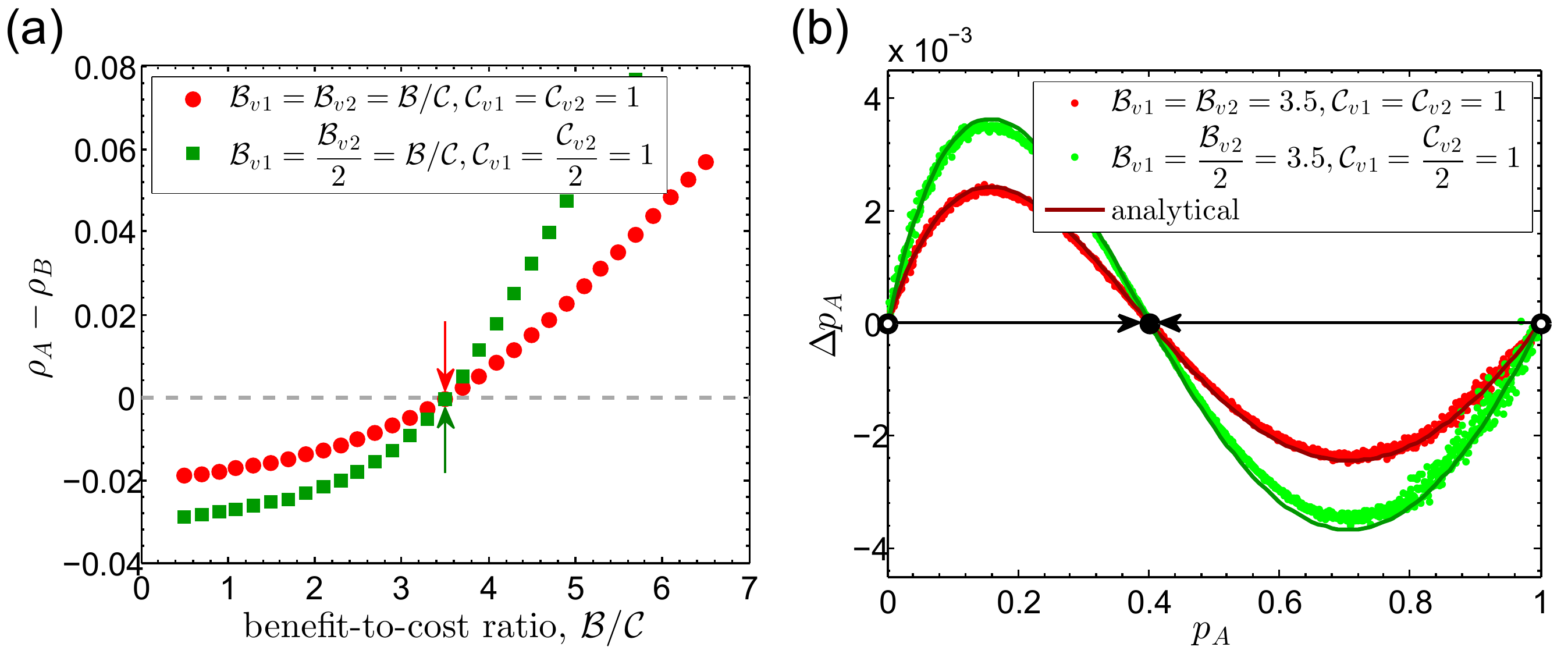}
\caption{\label{Fig 5} \textbf{Evolutionary multiplayer games on weighted graphs.}
Each individual participates in two volunteer's dilemmas and both group sizes are $4$.
Benefits and costs are $\mathcal{B}_{v1}$ and $\mathcal{C}_{v1}$ for one dilemma, $\mathcal{B}_{v2}$ and $\mathcal{C}_{v2}$ for the other.
(a) Difference in fixation probability $\rho_{\text{A}}-\rho_{\text{B}}$ as a function of benefit-to-cost ratio $\left(\mathcal{B}/\mathcal{C}\right)$.
(b) Average change ($\Delta p_{\text{A}}$) in the frequency of $A-$players ($p_{\text{A}}$).
Arrows in (a) mark the analytical benefit-to-cost ratio $\left(\mathcal{B}/\mathcal{C}\right)^*$ and solid lines in (b) represent analytical change in $p_{\text{A}}$.
Dots represent the simulation data (see Methods for simulation details).
Heterogeneous weights of edges do not change the critical benefit-to-cost ratio $\left(\mathcal{B}/\mathcal{C}\right)^*$ in the finite population (a) or the inner equilibria (black point) in the infinite population (b).
}
\end{figure*}

\renewcommand\thefigure{S1}

\begin{figure*}[!h]
\centering
\includegraphics[width=1\textwidth]{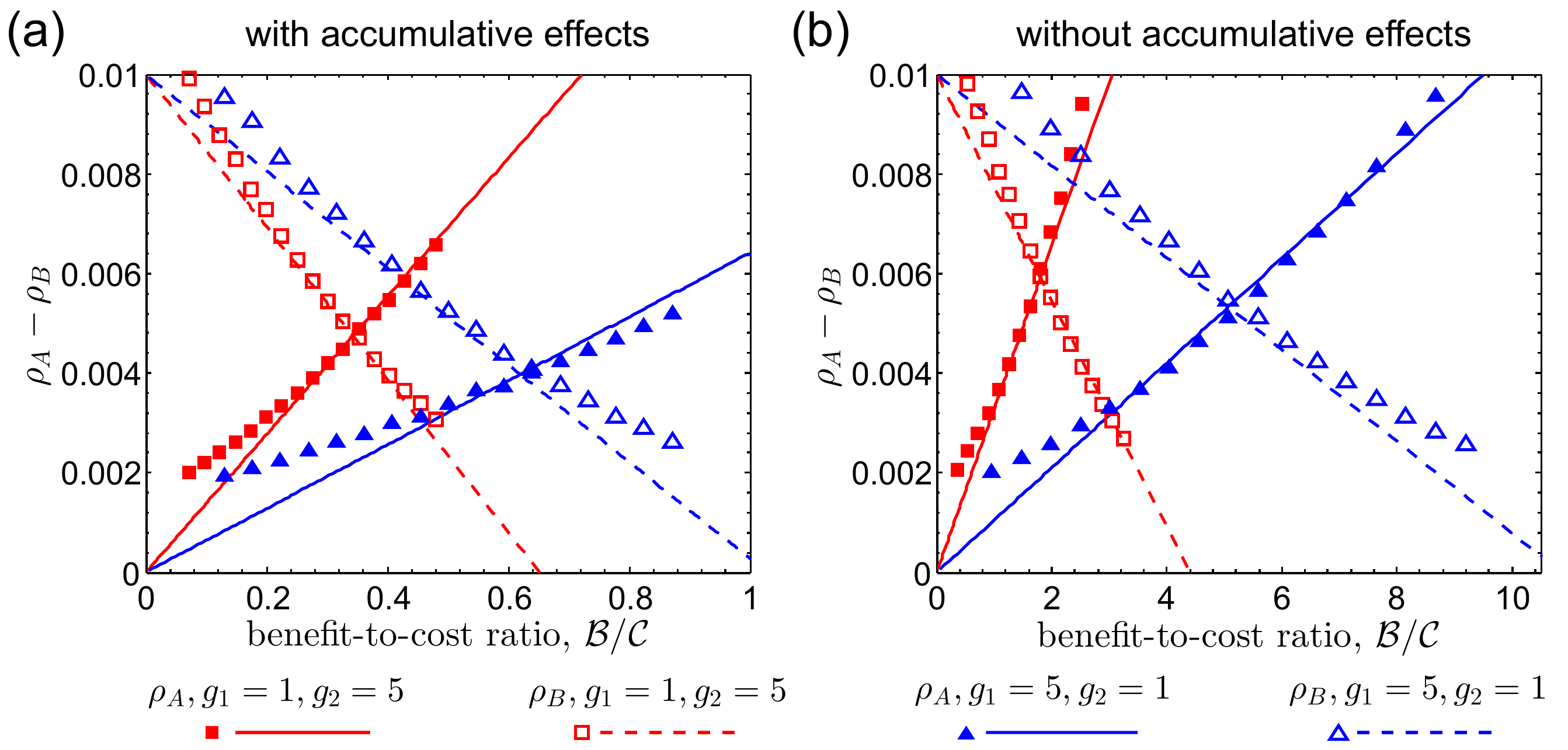}
\caption{\label{Fig 5} \textbf{Analytical fixation probability is in good agreement with simulation results.}
Solid lines present the analytical fixation probability of cooperators ($\rho_A$) and dash lines show the analytical fixation probability of defectors ($\rho_B$).
Dots show results by computer simulations.
Parameters in (a) follow Fig \ref{Fig 2}a and parameters in (b) follow Fig \ref{Fig 2}c.
}
\end{figure*}

\renewcommand\thefigure{S2}

\begin{figure*}[!h]
\centering
\includegraphics[width=1\textwidth]{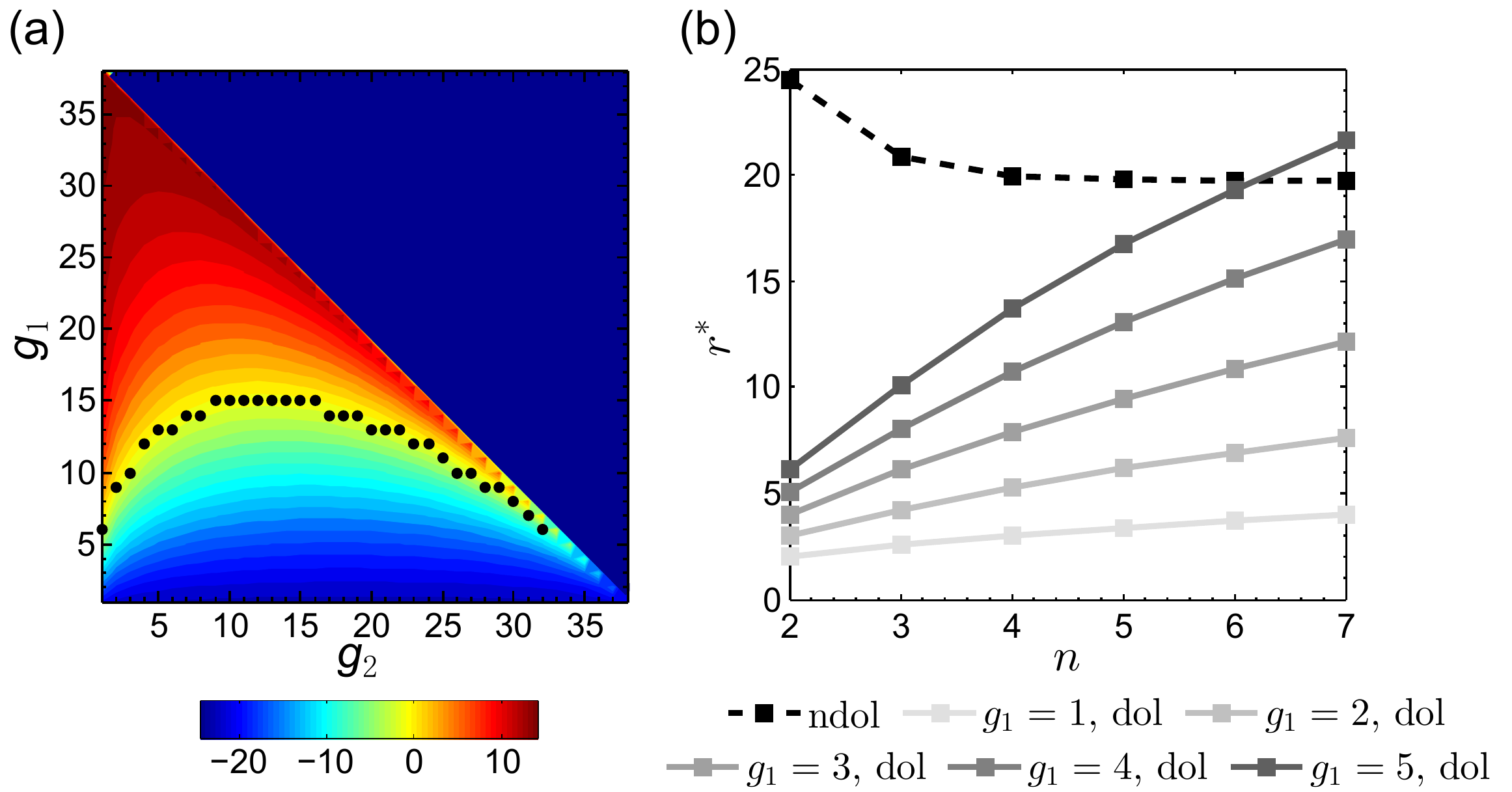}
\caption{\label{Fig 5} \textbf{Division of labor could reduce the free-riding behaviors for $n>2$.}
On graphs with $n$ types of edges, the production of benefits requires cooperation from players linked by each type of edges.
Note that the focal player and its neighbors linked by edges of type $1$ are of the same type.
Here the increasing number of cooperators does not lead to the increasing productivity, inasmuch as the number exceeds the threshold.
(a) Difference between $r^*$ with division of labor (``dol") and with no division of labor (``ndol").
$n=3$ and $g_1+g_2+g_3=40$.
The upper right zone is invalid given a positive $g_3$.
The block dots present the configurations of $g_1$ and $g_2$ for which $r^*$s with division of labor and with no division of labor are nearly equal.
(b) $r^*$ as a function of $n$.
We fix $\sum_{1\le i\le n}g_i=40$, $g_i=5$ for $2\le i\le n-1$, and vary $g_1$.
Both (a) and (b) show that a small value of $g_1$ facilitates cooperation.
}
\end{figure*}

\clearpage
\setcounter{section}{0}
\renewcommand{\thefigure}{S\arabic{figure}}
\textbf{SUPPLEMENTARY INFORMATION (SI)}

\section{Section 1. Fixation probability, structure coefficient, and replicator equation for evolutionary multiplayer games on graphs with $n$ types of edges}
\subsection{Pair approximation}
Let $p_A$ and $p_B$ be the frequencies of $A$-players and $B$-players in a population.
Let $p_{AA}$, $p_{AB}$, $p_{BA}$ and $p_{BB}$ be the frequencies of $AA$, $AB$, $BA$ and $BB$ pairs.
Let $q_{X|Y}$ be the conditional probability of finding an $X$-player given that the adjacent node is occupied by a $Y$-player, where $X$ and $Y$ are either $A$ or $B$.
Let $G_i$ denote the subgraph consisting of all nodes and edges of type $i$.
We distinguish aforementioned variables associated with $G_i$ labelling ($G_i$), such as $p_{AA}^{(G_1)}$ the frequencies of $AA$ pairs in $G_1$ and $p_{AA}^{(G_2)}$ in $G_2$.
$p_A^{(G_i)}$ is identical to $p_A^{(G_j)}$ for any pairs $i$, $j$, and thus we simplify them as $p_A$.
In the random regular graphs with $n$ types of edges,  we have identities
\begin{align}
p_A+p_B & = 1 \label{identity1}\\
p_{AB}^{(G_i)} & = p_{BA}^{(G_i)} \label{identity2}\\
q_{X|Y}^{(G_i)} & = \frac{p_{XY}^{(G_i)}}{p_Y} \label{identity3}\\
q_{A|Y}^{(G_i)}+q_{B|Y}^{(G_i)} & = 1 \label{identity4}
\end{align}
for any $1 \le i\le n$.
Eqs (\ref{identity1}-\ref{identity4}) imply that the whole system can be described by $n+1$ variables, i.e. $p_A$ and $q_{A|A}^{(G_i)}$ where $1 \le i\le n$.
These notations are given by
\begin{align}
p_B &= 1-p_A \nonumber \\
p_{AA}^{(G_i)} &= p_Aq_{A|A}^{(G_i)} \nonumber \\
p_{AB}^{(G_i)}=p_{BA}^{(G_i)} &= p_{A}(1-q_{A|A}^{(G_i)}) \nonumber \\
p_{BB}^{(G_i)} &= 1-2p_A+p_Aq_{A|A}^{(G_i)} \nonumber \\
q_{B|A}^{(G_i)} &= 1-q_{A|A}^{(G_i)} \nonumber \\
q_{B|B}^{(G_i)} &= \frac{1-2p_A+p_Aq_{A|A}^{(G_i)}}{1-p_A} \nonumber \\
q_{A|B}^{(G_i)} &= \frac{p_A(1-q_{A|A}^{(G_i)})}{1-p_A}  \nonumber
\end{align}
Let $g_i$ denote the node degree in $G_i$.
The node degree for the entire network is $k=\sum_{i=1}^n g_i$.
Let $a_{s_1\cdots s_n}$ be the payoff of an $A-$player that has $s_i$ neighboring $A-$players and $g_i-s_i$ neighboring $B-$players in $G_i$ ($1 \le i \le n$).
Let $b_{s_1\cdots s_n}$ be the payoff of a $B-$player that has $s_i$ neighboring $A-$players and $g_i-s_i$ neighboring $B-$players in $G_i$ ($1 \le i \le n$).
Each individual is assigned a payoff by a single interaction with all neighbors.
Then the payoff is transformed to the fitness for the evolution of system.
Here the population evolves based on the Moran death-birth process \cite{2006-Ohtsuki-p502-502}.
In each generation, a random individual is selected to die.
All neighbors compete to occupy the empty site proportional to their fitness.
Other process can be investigated analogously.
In the following, we calculate the expected change of $p_{A}$ and $p_{AA}^{(G_i)}$ in each step.

\subsection{Updating a B-player}
A $B$-player is selected to die with probability $p_B$.
Its $k$ neighbors compete to take over the vacant node.
Let $k_A^{(G_i)}$ and $k_B^{(G_i)}$ denote the number of $A$- and $B$-players among $g_i$ neighbors in $G_i$ ($1\le i\le n$).
We have $k_A^{(G_i)}+k_B^{(G_i)}=g_i$.
The probability for such a neighborhood configuration is
\begin{align} \label{probability_CPB}
\mathcal{B}_{k_A^{(G_1)}\cdots k_A^{(G_i)}\cdots k_A^{(G_n)}}^{g_1\cdots g_i\cdots g_n}=
\prod_{i=1}^n {g_i \choose k_A^{(G_i)}}(q_{A|B}^{(G_i)})^{k_A^{(G_i)}}(q_{B|B}^{(G_i)})^{k_B^{(G_i)}}.
\end{align}
Then the average fitness of each $A$-player and each $B$-player connected to this dead $B$-player by an edge in $G_i$ are respectively given by
\begin{align}
F_{A|B}^{(G_i)}=1-\omega+\omega\pi_{A|B}^{(G_i)}, \nonumber\\
F_{B|B}^{(G_i)}=1-\omega+\omega\pi_{B|B}^{(G_i)}, \nonumber
\end{align}
where
\begin{align}
\pi_{A|B}^{(G_i)}=\sum_{s_1=0}^{g_1}\cdots \sum_{s_i=0}^{g_i-1}\cdots \sum_{s_n=0}^{g_n}
\left[\prod_{j=1}^n {g_j-\delta_{i,j} \choose s_j}\left(q_{A|A}^{(G_j)}\right)^{s_j}\left(q_{B|A}^{(G_j)}\right)^{g_j-\delta_{i,j}-s_j}\right]
a_{s_1\cdots s_i\cdots s_n}, \label{fitnessAB1}\\
\pi_{B|B}^{(G_i)}=\sum_{s_1=0}^{g_1}\cdots \sum_{s_i=0}^{g_i-1}\cdots \sum_{s_n=0}^{g_n}
\left[\prod_{j=1}^n {g_j-\delta_{i,j} \choose s_j}\left(q_{A|B}^{(G_j)}\right)^{s_j}\left(q_{B|B}^{(G_j)}\right)^{g_j-\delta_{i,j}-s_j}\right]
b_{s_1\cdots s_i\cdots s_n} \label{fitnessBB1}
\end{align}
represent the expected payoffs from interactions with $\sum_{i=1}^n g_i$ neighbors.
$\delta_{i,j}=1$ if $j=i$ and $\delta_{i,j}=0$ if $j\ne i$.
The parameter $\omega$ denotes the intensity of selection and $w\ll 1$ means that the payoff from the game just contributes a little to one's fitness.
Here we consider the weak selection.

The probability that an $A$-player takes over the empty site is given by
\begin{align}
&\frac{\sum_{i=1}^n k_A^{(G_i)}F_{A|B}^{(G_i)}}{\sum_{i=1}^n k_A^{(G_i)}F_{A|B}^{(G_i)}+\sum_{i=1}^n k_B^{(G_i)}F_{B|B}^{(G_i)}} \nonumber \\
=&\frac{\sum_{i=1}^n k_A^{(G_i)}}{k}+\frac{\omega}{k^2}\sum_{i=1}^n\sum_{j=1}^n k_A^{(G_i)}k_B^{(G_j)}\left(\pi_{A|B}^{(G_i)}-\pi_{B|B}^{(G_j)}\right)+O(\omega^2). \label{probability_AB}
\end{align}

Therefore, combining Eqs~(\ref{probability_CPB},\ref{probability_AB}), $p_A$ increases by $\frac{1}{N}$ with probability
\begin{align}
\text{Prob}\left(\Delta p_A=\frac{1}{N}\right)
=& p_A\sum_{k_A^{(G_1)}=0}^{g_1}\cdots \sum_{k_A^{(G_i)}=0}^{g_i}\cdots \sum_{k_A^{(G_n)}=0}^{g_n}
\mathcal{B}_{k_A^{(G_1)}\cdots k_A^{(G_i)}\cdots k_A^{(G_n)}}^{g_1\cdots g_i\cdots g_n} \nonumber \\
&\frac{\sum_{i=1}^n k_A^{(G_i)}F_{A|B}^{(G_i)}}{\sum_{i=1}^n k_A^{(G_i)}F_{A|B}^{(G_i)}+\sum_{i=1}^n k_B^{(G_i)}F_{B|B}^{(G_i)}} \nonumber \\
=&\frac{p_B}{k}\sum_{i=1}^ng_i q_{A|B}^{(G_i)}+\frac{\omega p_B}{k^2}\Gamma_B+O(\omega^2)\nonumber
\end{align}
where
\begin{align} \label{gammaB}
\Gamma_B =
& \sum_{i=1}^n\sum_{j=1}^n g_ig_jq_{A|B}^{(G_i)}q_{B|B}^{(G_j)}\left(\pi_{A|B}^{(G_i)}-\pi_{B|B}^{(G_j)}\right)
-\sum_{i=1}^n g_iq_{A|B}^{(G_i)}q_{B|B}^{(G_i)}\left(\pi_{A|B}^{(G_i)}-\pi_{B|B}^{(G_i)}\right)
\end{align}

Regarding pairs, if an $A$-player in $G_i$ occupies the vacant site then the number of $AA$ pairs in $G_i$ increases by
$k_A^{(G_i)}$.
Given that the total number of pairs in $G_i$ is $g_iN/2$, the probability that $p_{AA}^{(G_i)}$ increases by $2k_A^{(G_i)}/(g_iN)$ is given by
\begin{align}
\text{Prob}\left(\Delta p_{AA}^{(G_i)}=\frac{2k_A^{(G_i)}}{g_iN}\right)
=&p_B\sum_{k_A^{(G_1)}=0}^{g_1}\cdots \sum_{k_A^{(G_{i-1})}=0}^{g_{i-1}}\sum_{k_A^{(G_{i+1})}=0}^{g_{i+1}}\cdots \sum_{k_A^{(G_n)}=0}^{g_n}\mathcal{B}_{k_A^{(G_1)}\cdots k_A^{(G_i)}\cdots k_A^{(G_n)}}^{g_1\cdots g_i\cdots g_n} \nonumber\\
&\frac{\sum_{j=1}^n k_A^{(G_j)}F_{A|B}^{(G_j)}}{\sum_{j=1}^n k_A^{(G_j)}F_{A|B}^{(G_j)}+\sum_{j=1}^n k_B^{(G_j)}F_{B|B}^{(G_j)}} \nonumber
\end{align}

\subsection{Updating an $A$-player}
An $A$-player is selected to die with probability $p_A$.
All $k$ individuals, i.e., $k_A^{(G_i)}$ $A$-players and $k_B^{(G_i)}$ $B$-players in $G_i$ ($1\le i\le n$), compete to occupy the empty site.
The probability for such a neighborhood configuration is given by
\begin{align} \label{probability_DPA}
\mathcal{A}_{k_A^{(G_1)}\cdots k_A^{(G_i)}\cdots k_A^{(G_n)}}^{g_1\cdots g_i\cdots g_n}=
\prod_{i=1}^n {g_i \choose k_A^{(G_i)}}(q_{A|A}^{(G_i)})^{k_A^{(G_i)}}(q_{B|A}^{(G_i)})^{k_B^{(G_i)}}.
\end{align}
Then the average fitness of each $A$-player and each $B$-player connected to this dead $A$-player by an edge in $G_i$ are respectively given by
\begin{align}
F_{A|A}^{(G_i)}=1-\omega+\omega\pi_{A|A}^{(G_i)}, \nonumber\\
F_{B|A}^{(G_i)}=1-\omega+\omega\pi_{B|A}^{(G_i)}, \nonumber
\end{align}
where
\begin{align}
\pi_{A|A}^{(G_i)}=\sum_{s_1=0}^{g_1}\cdots \sum_{s_i=0}^{g_i-1}\cdots \sum_{s_n=0}^{g_n}
\left[\prod_{j=1}^n {g_j-\delta_{i,j} \choose s_j}\left(q_{A|A}^{(G_j)}\right)^{s_j}\left(q_{B|A}^{(G_j)}\right)^{g_j-\delta_{i,j}-s_j}\right]
a_{s_1\cdots (s_i+1)\cdots s_n}, \label{fitnessAA1}\\
\pi_{B|A}^{(G_i)}=\sum_{s_1=0}^{g_1}\cdots \sum_{s_i=0}^{g_i-1}\cdots \sum_{s_n=0}^{g_n}
\left[\prod_{j=1}^n {g_j-\delta_{i,j} \choose s_j}\left(q_{A|B}^{(G_j)}\right)^{s_j}\left(q_{B|B}^{(G_j)}\right)^{g_j-\delta_{i,j}-s_j}\right]
b_{s_1\cdots (s_i+1)\cdots s_n} \label{fitnessBA1}
\end{align}
represent the expected payoffs from interactions with $\sum_{i=1}^n g_i$ neighbors.

The probability that a $B$-player takes over the empty site with probability
\begin{align}
&\frac{\sum_{i=1}^n k_B^{(G_i)}F_{B|A}^{(G_i)}}{\sum_{i=1}^n k_A^{(G_i)}F_{A|A}^{(G_i)}+\sum_{i=1}^n k_B^{(G_i)}F_{B|A}^{(G_i)}} \nonumber \\
=&\frac{\sum_{i=1}^n k_B^{(G_i)}}{k}+\frac{\omega}{k^2}
\sum_{i=1}^n\sum_{j=1}^n k_A^{(G_i)}k_B^{(G_j)}\left(\pi_{B|A}^{(G_j)}-\pi_{A|A}^{(G_i)}\right)+O(\omega^2). \label{probability_BA}
\end{align}

Therefore, combining Eqs~(\ref{probability_DPA}) and (\ref{probability_BA}), $p_A$ decreases by $\frac{1}{N}$ with probability
\begin{align}
\text{Prob}\left(\Delta p_A=-\frac{1}{N}\right)
=& p_A\sum_{k_A^{(G_1)}=0}^{g_1}\cdots \sum_{k_A^{(G_i)}=0}^{g_i}\cdots \sum_{k_A^{(G_n)}=0}^{g_n}
\mathcal{A}_{k_A^{(G_1)}\cdots k_A^{(G_i)}\cdots k_A^{(G_n)}}^{g_1\cdots g_i\cdots g_n} \nonumber \\
&\frac{\sum_{i=1}^n k_B^{(G_i)}F_{B|A}^{(G_i)}}{\sum_{i=1}^n k_A^{(G_i)}F_{A|A}^{(G_i)}+\sum_{i=1}^n k_B^{(G_i)}F_{B|A}^{(G_i)}} \nonumber \\
=&\frac{p_A}{k}\sum_{i=1}^ng_i q_{B|A}^{(G_i)}+\frac{\omega p_A}{k^2}\Gamma_A+O(\omega^2)\nonumber
\end{align}
where
\begin{align} \label{gammaA}
\Gamma_A =
& \sum_{i=1}^n\sum_{j=1}^n g_ig_jq_{A|A}^{(G_i)}q_{B|A}^{(G_j)}\left(\pi_{B|A}^{(G_j)}-\pi_{A|A}^{(G_i)}\right)
-\sum_{i=1}^n g_iq_{A|A}^{(G_i)}q_{B|A}^{(G_i)}\left(\pi_{B|A}^{(G_i)}-\pi_{A|A}^{(G_i)}\right)
\end{align}

Regarding pairs, if a $B$-player in $G_i$ occupies the vacant site then the number of $AA$ pairs in $G_i$ decreases by
$k_A^{(G_i)}$ and therefore $p_{AA}^{(G_i)}$ decreases by $2k_A^{(G_i)}/(g_iN)$ with probability
\begin{align}
\text{Prob}\left(\Delta p_{AA}^{(G_i)}=-\frac{2k_A^{(G_i)}}{g_iN}\right)
=&p_A\sum_{k_A^{(G_1)}=0}^{g_1}\cdots \sum_{k_A^{(G_{i-1})}=0}^{g_{i-1}}\sum_{k_A^{(G_{i+1})}=0}^{g_{i+1}}\cdots \sum_{k_A^{(G_n)}=0}^{g_n}\mathcal{A}_{k_A^{(G_1)}\cdots k_A^{(G_i)}\cdots k_A^{(G_n)}}^{g_1\cdots g_i\cdots g_n} \nonumber\\
&\frac{\sum_{j=1}^n k_B^{(G_j)}F_{B|A}^{(G_j)}}{\sum_{j=1}^n k_A^{(G_j)}F_{A|A}^{(G_j)}+\sum_{j=1}^n k_B^{(G_j)}F_{B|A}^{(G_j)}} \nonumber
\end{align}

\subsection{Different time scales}
Supposing that one replacement event takes place in one unit of time, we can get the time derivatives of $p_A$ and $p_{AA}^{(G_i)}$, given by
\begin{align} \label{derivative_pA}
\dot{p}_A
=& \frac{1}{N}\cdot\text{Prob}\left(\Delta p_A=\frac{1}{N}\right)+\left(-\frac{1}{N}\right)\cdot\text{Prob}\left(\Delta p_A=-\frac{1}{N}\right) \nonumber\\
=& \frac{\omega}{Nk^2}\left(p_B\Gamma_B-p_A\Gamma_A\right)+O(\omega^2),
\end{align}
\begin{align} \label{derivative_pAAG1}
\dot{p}_{AA}^{(G_i)}
=& \sum_{k_A^{(G_i)}=0}^{g_i}\frac{2k_A^{(G_i)}}{g_iN}\cdot\text{Prob}\left(\Delta p_{AA}^{(G_i)}=\frac{2k_A^{(G_i)}}{g_iN}\right)+
\sum_{k_A^{(G_i)}=0}^{g_i}\left(-\frac{2k_A^{(G_i)}}{g_iN}\right)\cdot\text{Prob}\left(\Delta p_{AA}^{(G_i)}=-\frac{2k_A^{(G_i)}}{g_iN}\right) \nonumber\\
=& \frac{2p_A}{Nk(1-p_A)}\left[
\sum_{j=1}^ng_j\left(1-q_{A|A}^{(G_j)}\right)\left(p_A-q_{A|A}^{(G_i)}\right)+\left(1-q_{A|A}^{(G_i)}\right)\left(1+q_{A|A}^{(G_i)}-2p_A\right)
\right]+O(\omega),
\end{align}
From Eqs (\ref{derivative_pA}) and (\ref{derivative_pAAG1}), we have
\begin{align} \label{derivative_qAAG1}
\dot{q}_{A|A}^{(G_i)}
& =\frac{\text{d}}{\text{d}t}\left(\frac{p_{AA}^{(G_i)}}{p_A}\right) \nonumber \\
& =\frac{2}{Nk(1-p_A)}\left[
\sum_{j=1}^ng_j\left(1-q_{A|A}^{(G_j)}\right)\left(p_A-q_{A|A}^{(G_i)}\right)+\left(1-q_{A|A}^{(G_i)}\right)\left(1+q_{A|A}^{(G_i)}-2p_A\right)
\right]+O(\omega).
\end{align}
Rewriting Eqs~(\ref{derivative_pA}) and (\ref{derivative_qAAG1}) as a function of $p_A$ and $q_{A|A}^{(G_i)}$, we have
\begin{align}
\dot{p}_A &= \omega\cdot \Psi_0(p_A,q_{A|A}^{(G_1)},\cdots,q_{A|A}^{(G_n)})+O(\omega^2), \nonumber \\
\dot{q}_{A|A}^{(G_i)} &= \Psi_i(p_A,q_{A|A}^{(G_1)},\cdots,q_{A|A}^{(G_n)})+O(\omega). \nonumber
\end{align}
For weak selection ($\omega\ll 1$), $q_{A|A}^{(G_i)}$ equilibrates much more quickly than $p_A$.
Thus, this dynamical system converges rapidly onto the slow manifold with $\Psi_i(p_A,q_{A|A}^{(G_1)},\cdots,q_{A|A}^{(G_n)})=0$.
Then we have
\begin{align}
q_{A|A}^{(G_i)}&=\frac{\sum_{j=1}^ng_j-2}{\sum_{j=1}^ng_j-1}p_A+\frac{1}{\sum_{j=1}^ng_j-1}=\frac{k-2}{k-1}p_A+\frac{1}{k-1}. \label{scale_qAAG1}
\end{align}

Defining $z=\frac{1}{k-1}$ and using Eq~(\ref{scale_qAAG1}), we rewrite $q_{X|Y}^{(G_i)}$ as
\begin{align}
q_{A|A}^{(G_i)}&=p_A+z(1-p_A), \label{qAAz} \\
q_{B|A}^{(G_i)}&=(1-z)(1-p_A), \label{qBAz} \\
q_{A|B}^{(G_i)}&=(1-z)p_A, \label{qABz} \\
q_{B|B}^{(G_i)}&=zp_A+(1-p_A). \label{qBBz}
\end{align}

\subsection{Diffusion approximation}
Equation (\ref{scale_qAAG1}) holds for all $G_i$.
We use Kolmogorov backward equation to study an one dimensional diffusion process of variable $p_A$.
The fixation probability of $A$-players, $\phi_A(x)$ with initial frequency $p_A(t=0)=x$ satisfies the differential equation \cite{2004-Gardiner}:
\begin{align}
m(x)\frac{d\phi_A(x)}{dx}+\frac{v(x)}{2}\frac{d^2\phi_A(x)}{dx^2}=0
\end{align}
with two boundary conditions $\phi_A(0)=0$ and $\phi_A(1)=1$.
$m(p_A)$ and $v(p_A)$ represent the mean and variance of $\Delta p_A$ in each generation, respectively.
The solution for the above differential equation with boundary conditions is
\begin{align}
\phi_A(x)=\frac{\int_{0}^{x}\psi(y)dy}{\int_{0}^{1}\psi(y)dy} \label{solution1}
\end{align}
where
\begin{align}
\psi(y)=\text{exp}\left(-\int^y \frac{2m(r)}{v(r)}dr\right). \label{solution2}
\end{align}

Within a short time interval, $\Delta t$, we have
\begin{align} \label{mpA}
m(p_A)
&=\frac{\text{E}(\Delta p_A)}{\Delta t} \nonumber \\
&=\frac{1}{N}\cdot\text{Prob}\left(\Delta p_A=\frac{1}{N}\right)+\left(-\frac{1}{N}\right)\cdot\text{Prob}\left(\Delta p_A=-\frac{1}{N}\right)   \nonumber \\
&\approx \frac{\omega}{Nk^2}\left(p_B\Gamma_B-p_A\Gamma_A\right)
\end{align}
\begin{align} \label{vpA}
v(p_A)
&=\frac{\text{V}(\Delta p_A)}{\Delta t} \nonumber \\
&\approx \frac{1}{N^2}\cdot\text{Prob}\left(\Delta p_A=\frac{1}{N}\right)+\frac{1}{N^2}\cdot\text{Prob}\left(\Delta p_A=-\frac{1}{N}\right) \nonumber \\
&\approx \frac{2(k-2)p_A(1-p_A)}{N^2(k-1)}
\end{align}
According to Eqs~(\ref{fitnessAB1},\ref{fitnessBB1},\ref{gammaB},\ref{fitnessAA1},\ref{fitnessBA1},\ref{gammaA},\ref{qAAz}-\ref{qBBz}), $p_B\Gamma_B-p_A\Gamma_A$ in Eq~(\ref{mpA}) is actually a polynomial in $p_A$.
We here can write it in a form only containing variable $p_A$.
To achieve this, we make full use of Eqs~(19-22) in Supplemental Methods of Ref.~\cite{2016-Pena-p1-15} and Appendix B of Ref.~\cite{2015-Pena-p122-136} to obtain the following identities:
\begin{align}
&\sum_{s_1=0}^{g_1}\cdots \sum_{s_i=0}^{g_i}\cdots \sum_{s_n=0}^{g_n}
\left[\prod_{j=1}^n {g_j \choose s_j}\left[x+z(1-x)\right]^{s_j}\left[(1-z)(1-x)\right]^{g_j-s_j}\right] a_{s_1\cdots s_i\cdots s_n} \nonumber \\
=&\sum_{s_1=0}^{g_1}\cdots \sum_{s_i=0}^{g_i}\cdots \sum_{s_n=0}^{g_n}
\left[\prod_{j=1}^n {g_j \choose s_j}x^{s_j}(1-x)^{g_j-s_j}\right] \nonumber \\
&\sum_{r_1=0}^{g_1-s_1}\cdots \sum_{r_i=0}^{g_i-s_i}\cdots \sum_{r_n=0}^{g_n-s_n}
\left[\prod_{j=1}^n {g_j-s_j \choose r_j}z^{r_j}(1-z)^{g_j-s_j-r_j}\right]
a_{(s_1+r_1)\cdots (s_i+r_i)\cdots (s_n+r_n)}, \label{identity_a1} \\
&\sum_{s_1=0}^{g_1}\cdots \sum_{s_i=0}^{g_i}\cdots \sum_{s_n=0}^{g_n}
\left[\prod_{j=1}^n {g_j \choose s_j}\left[(1-z)x\right]^{s_j}\left[zx+1-x\right]^{g_j-s_j}\right] a_{s_1\cdots s_i\cdots s_n} \nonumber \\
=&\sum_{s_1=0}^{g_1}\cdots \sum_{s_i=0}^{g_i}\cdots \sum_{s_n=0}^{g_n}
\left[\prod_{j=1}^n {g_j \choose s_j}x^{s_j}(1-x)^{g_j-s_j}\right] \nonumber \\
&\sum_{r_1=0}^{s_1}\cdots \sum_{r_i=0}^{s_i}\cdots \sum_{r_n=0}^{s_n}
\left[\prod_{j=1}^n {s_j \choose r_j}z^{r_j}(1-z)^{s_j-r_j}\right]
a_{(s_1-r_1)\cdots (s_i-r_i)\cdots (s_n-r_n)}, \label{identity_a2} \\
%\sum_{j=0}^{h}{h \choose j}\left[x+z(1-x)\right]^j\left[(1-z)(1-x)\right]^{h-j}a_{j}^{i} \nonumber\\
%=& \sum_{i=0}^{g}\sum_{j=0}^{h}{g \choose i}{h \choose j}x^{i+j}(1-x)^{g+h-i-j}
%\sum_{m=0}^{g-i}\sum_{n=0}^{h-j}{g-i \choose m}{h-j \choose n}z^{m+n}(1-z)^{g+h-i-j-m-n}a_{j+n}^{i+m}, \label{identity_a1}\\
%&\sum_{i=0}^{g}{g \choose i}\left[(1-z)x\right]^i\left[zx+1-x\right]^{g-i}
%\sum_{j=0}^{h}{h \choose j}\left[(1-z)x\right]^j\left[zx+1-x\right]^{h-j}a_{j}^{i} \nonumber\\
%=& \sum_{i=0}^{g}\sum_{j=0}^{h}{g \choose i}{h \choose j}x^{i+j}(1-x)^{g+h-i-j}
%\sum_{m=0}^{i}\sum_{n=0}^{j}{i \choose m}{j \choose n}z^{m+n}(1-z)^{i+j-m-n}a_{j-n}^{i-m}, \label{identity_a2}\\
%& x\sum_{i=0}^{g-1}\sum_{j=0}^{h}{g-1 \choose i}{h \choose j}x^{i+j}(1-x)^{g+h-1-i-j}a_{j}^{i}
%=\sum_{i=0}^{g}\sum_{j=0}^{h}{g \choose i}{h \choose j}x^{i+j}(1-x)^{g+h-i-j}\frac{ia_{j}^{i-1}}{g}, \label{identity_a3}
&x\sum_{s_1=0}^{g_1}\cdots \sum_{s_i=0}^{g_i-1}\cdots \sum_{s_n=0}^{g_n}\left[\prod_{j=1}^n {g_j-\delta_{i,j} \choose s_j}x^{s_j}(1-x)^{g_j-\delta_{i,j}-s_j}\right]a_{s_1\cdots s_i\cdots s_n} \nonumber \\
&=\sum_{s_1=0}^{g_1}\cdots \sum_{s_i=0}^{g_i}\cdots \sum_{s_n=0}^{g_n}
\left[\prod_{j=1}^n {g_j \choose s_j}x^{s_j}(1-x)^{g_j-s_j}\right]\frac{s_ia_{s_1\cdots (s_i-1)\cdots s_n}}{g_i}, \label{identity_a3} \\
&(1-x)\sum_{s_1=0}^{g_1}\cdots \sum_{s_i=0}^{g_i-1}\cdots \sum_{s_n=0}^{g_n}\left[\prod_{j=1}^n {g_j-\delta_{i,j} \choose s_j}x^{s_j}(1-x)^{g_j-\delta_{i,j}-s_j}\right]a_{s_1\cdots s_i\cdots s_n} \nonumber \\
&=\sum_{s_1=0}^{g_1}\cdots \sum_{s_i=0}^{g_i}\cdots \sum_{s_n=0}^{g_n}
\left[\prod_{j=1}^n {g_j \choose s_j}x^{s_j}(1-x)^{g_j-s_j}\right]\frac{(g_i-s_i)a_{s_1\cdots s_i\cdots s_n}}{g_i}. \label{identity_a4}
\end{align}

Replacing Eqs~(\ref{qAAz}-\ref{qBBz}) into Eqs~(\ref{fitnessAB1},\ref{fitnessBB1},\ref{fitnessAA1},\ref{fitnessBA1}) and applying Eqs~(\ref{identity_a1}) and (\ref{identity_a2}), we have
\begin{align}
\pi_{A|B}^{(G_i)}=
&\sum_{s_1=0}^{g_1}\cdots \sum_{s_i=0}^{g_i-1}\cdots \sum_{s_n=0}^{g_n}
\left[\prod_{j=1}^n {g_j-\delta_{i,j} \choose s_j}p_A^{s_j}(1-p_A)^{g_j-\delta_{i,j}-s_j}\right] \nonumber \\
&\sum_{r_1=0}^{g_1-s_1}\cdots \sum_{r_i=0}^{g_i-1-s_i}\cdots \sum_{r_n=0}^{g_n-s_n}
\left[\prod_{j=1}^n {g_j-\delta_{i,j}-s_j \choose r_j}z^{r_j}(1-z)^{g_j-\delta_{i,j}-s_j-r_j}\right]
a_{(s_1+r_1)\cdots (s_i+r_i)\cdots (s_n+r_n)}, \label{piABG1_new}
\end{align}
%%%
\begin{align}
\pi_{B|B}^{(G_i)}=
&\sum_{s_1=0}^{g_1}\cdots \sum_{s_i=0}^{g_i-1}\cdots \sum_{s_n=0}^{g_n}
\left[\prod_{j=1}^n {g_j-\delta_{i,j} \choose s_j}p_A^{s_j}(1-p_A)^{g_j-\delta_{i,j}-s_j}\right] \nonumber \\
&\sum_{r_1=0}^{s_1}\cdots \sum_{r_i=0}^{s_i}\cdots \sum_{r_n=0}^{s_n}
\left[\prod_{j=1}^n {s_j \choose r_j}z^{r_j}(1-z)^{s_j-r_j}\right]
b_{(s_1-r_1)\cdots (s_i-r_i)\cdots (s_n-r_n)}, \label{piBBG1_new}\\
%%%
%%%%
\pi_{A|A}^{(G_i)}=
&\sum_{s_1=0}^{g_1}\cdots \sum_{s_i=0}^{g_i-1}\cdots \sum_{s_n=0}^{g_n}
\left[\prod_{j=1}^n {g_j-\delta_{i,j} \choose s_j}p_A^{s_j}(1-p_A)^{g_j-\delta_{i,j}-s_j}\right] \nonumber \\
&\sum_{r_1=0}^{g_1-s_1}\cdots \sum_{r_i=0}^{g_i-1-s_i}\cdots \sum_{r_n=0}^{g_n-s_n}
\left[\prod_{j=1}^n {g_j-\delta_{i,j}-s_j \choose r_j}z^{r_j}(1-z)^{g_j-\delta_{i,j}-s_j-r_j}\right]
a_{(s_1+r_1)\cdots (s_i+r_i+1)\cdots (s_n+r_n)} \label{piABG1_newA}, \\
%%%
\pi_{B|A}^{(G_i)}=
&\sum_{s_1=0}^{g_1}\cdots \sum_{s_i=0}^{g_i-1}\cdots \sum_{s_n=0}^{g_n}
\left[\prod_{j=1}^n {g_j-\delta_{i,j} \choose s_j}p_A^{s_j}(1-p_A)^{g_j-\delta_{i,j}-s_j}\right] \nonumber \\
&\sum_{r_1=0}^{s_1}\cdots \sum_{r_i=0}^{s_i}\cdots \sum_{r_n=0}^{s_n}
\left[\prod_{j=1}^n {s_j \choose r_j}z^{r_j}(1-z)^{s_j-r_j}\right]
b_{(s_1-r_1)\cdots (s_i-r_i+1)\cdots (s_n-r_n)}, \label{piBBG1_newA}
\end{align}

Substituting Eqs~(\ref{piABG1_new}-\ref{piBBG1_newA}) into Eq~(\ref{mpA}) and applying Eqs~(\ref{identity_a3}) and (\ref{identity_a4}), we have
\begin{align}
m(p_A)
\approx &\frac{\omega(k-2)p_A(1-p_A)}{Nk^2}\sum_{i=1}^n \left[g_iq_{B|B}^{(G_i)}\left(\pi_{A|B}^{(G_i)}-\pi_{B|B}^{(G_i)}\right)+g_iq_{A|A}^{(G_i)}\left(\pi_{A|A}^{(G_i)}-\pi_{B|A}^{(G_i)}\right)\right] \nonumber \\
=&\frac{\omega(k-2)p_A(1-p_A)}{Nk^2}\sum_{i=1}^ng_i \nonumber\\
&\Bigg\{\left[zp_A+(1-p_A)\right]\sum_{s_1=0}^{g_1}\cdots \sum_{s_i=0}^{g_i-1}\cdots \sum_{s_n=0}^{g_n}
\left[\prod_{j=1}^n {g_j-\delta_{i,j} \choose s_j}p_A^{s_j}(1-p_A)^{g_j-\delta_{i,j}-s_j}\right]\leftidx{^i}c_{s_1\cdots s_i\cdots s_n}
\nonumber \\
&+\left[p_A+z(1-p_A)\right]\sum_{s_1=0}^{g_1}\cdots \sum_{s_i=0}^{g_i-1}\cdots \sum_{s_n=0}^{g_n}
\left[\prod_{j=1}^n {g_j-\delta_{i,j} \choose s_j}p_A^{s_j}(1-p_A)^{g_j-\delta_{i,j}-s_j}\right]\leftidx{^i}d_{s_1\cdots s_i\cdots s_n} \Bigg\} \nonumber \\
=&\frac{\omega(k-2)p_A(1-p_A)}{Nk^2}\sum_{s_1=0}^{g_1}\cdots \sum_{s_i=0}^{g_i}\cdots \sum_{s_n=0}^{g_n}
\left[\prod_{j=1}^n {g_j \choose s_j}p_A^{s_j}(1-p_A)^{g_j-s_j}\right]e_{s_1\cdots s_i\cdots s_n} \label{mpA2}
\end{align}
where
\begin{align} \label{eji}
e_{s_1\cdots s_i\cdots s_n}=\sum_{i=1}^n \left[zs_i\leftidx{^i}c_{s_1\cdots (s_i-1)\cdots s_n}+(g_i-s_i)\leftidx{^i}c_{s_1\cdots s_i\cdots s_n}+s_i\leftidx{^i}d_{s_1\cdots (s_i-1)\cdots s_n}+z(g_i-s_i)\leftidx{^i}d_{s_1\cdots s_i\cdots s_n}\right]
\end{align}
and
\begin{align}
&\leftidx{^i}c_{s_1\cdots s_i\cdots s_n} \nonumber \\=
&\sum_{r_1=0}^{g_1-s_1}\cdots \sum_{r_i=0}^{g_i-1-s_i}\cdots \sum_{r_n=0}^{g_n-s_n}
\left[\prod_{j=1}^n {g_j-\delta_{i,j}-s_j \choose r_j}z^{r_j}(1-z)^{g_j-\delta_{i,j}-s_j-r_j}\right]
a_{(s_1+r_1)\cdots (s_i+r_i)\cdots (s_n+r_n)} \nonumber\\
&-\sum_{r_1=0}^{s_1}\cdots \sum_{r_i=0}^{s_i}\cdots \sum_{r_n=0}^{s_n}
\left[\prod_{j=1}^n {s_j \choose r_j}z^{r_j}(1-z)^{s_j-r_j}\right]
b_{(s_1-r_1)\cdots (s_i-r_i)\cdots (s_n-r_n)} \label{cji}\\
&\leftidx{^i}d_{s_1\cdots s_i\cdots s_n} \nonumber \\=
&\sum_{r_1=0}^{g_1-s_1}\cdots \sum_{r_i=0}^{g_i-1-s_i}\cdots \sum_{r_n=0}^{g_n-s_n}
\left[\prod_{j=1}^n {g_j-\delta_{i,j}-s_j \choose r_j}z^{r_j}(1-z)^{g_j-\delta_{i,j}-s_j-r_j}\right]
a_{(s_1+r_1)\cdots (s_i+r_i+1)\cdots (s_n+r_n)} \nonumber\\
&-\sum_{r_1=0}^{s_1}\cdots \sum_{r_i=0}^{s_i}\cdots \sum_{r_n=0}^{s_n}
\left[\prod_{j=1}^n {s_j \choose r_j}z^{r_j}(1-z)^{s_j-r_j}\right]
b_{(s_1-r_1)\cdots (s_i-r_i+1)\cdots (s_n-r_n)}. \label{dji}
\end{align}

Denoting
\begin{align}
H(p_A)=\sum_{s_1=0}^{g_1}\cdots \sum_{s_i=0}^{g_i}\cdots \sum_{s_n=0}^{g_n}\left[\prod_{j=1}^n {g_j \choose s_j}p_A^{s_j}(1-p_A)^{g_j-s_j}\right]e_{s_1\cdots s_i\cdots s_n}
\end{align}
and substituting Eqs~(\ref{vpA}) and (\ref{mpA2}) into
Eqs~(\ref{solution1}) and (\ref{solution2}), for $\omega\ll 1$, we have
\begin{align}
\phi_A(x)=
&\frac{\int_0^x \text{exp}\left(-\frac{\omega N(k-1)}{k^2}\int_0^y H(r)dr\right)dy}
{\int_0^1 \text{exp}\left(-\frac{\omega N(k-1)}{k^2}\int_0^y H(r)dr\right)dy}       \nonumber \\
=&x+\frac{\omega N(k-1)}{k^2}\left(x\int_0^1\int_0^yH(r)drdy-\int_0^x\int_0^yH(r)drdy\right)+O(\omega^2)
\end{align}
Based on the integral property of Bernstein polynomial \cite{2012-Farouki-p379-419},
\begin{align}
\int_0^y {g \choose i}r^i(1-r)^{g-i}dr = \frac{1}{g+1}\sum_{j=i+1}^{g+1}{g+1 \choose j}y^j(1-y)^{g+1-j}
\end{align}
we have
\begin{align} \label{interal1}
\int_0^x\int_0^yH(r)drdy
=&\int_0^x\int_0^y \sum_{s_1=0}^{g_1}\cdots \sum_{s_i=0}^{g_i}\cdots \sum_{s_n=0}^{g_n}\left[\prod_{j=1}^n {g_j \choose s_j}r^{s_j}(1-r)^{g_j-s_j}\right]e_{s_1\cdots s_i\cdots s_n} drdy \nonumber\\
=&\int_0^x \sum_{s_1=0}^{g_1}\cdots \sum_{s_i=0}^{g_i}\cdots \sum_{s_n=0}^{g_n} \frac{\prod_{j=1}^n {g_j \choose s_j}}{{\sum_{j=1}^n g_j \choose \sum_{j=1}^n s_j}}e_{s_1\cdots s_i\cdots s_n} \nonumber\\
&\int_0^y {\sum_{j=1}^n g_j \choose \sum_{j=1}^n s_j}r^{\sum_{j=1}^n s_j}(1-r)^{\sum_{j=1}^n g_j-\sum_{j=1}^n s_j}drdy \nonumber \\
=&\frac{1}{\sum_{j=1}^n g_j+1}\int_0^x \sum_{s_1=0}^{g_1}\cdots \sum_{s_i=0}^{g_i}\cdots \sum_{s_n=0}^{g_n} \frac{\prod_{j=1}^n {g_j \choose s_j}}{{\sum_{j=1}^n g_j \choose \sum_{j=1}^n s_j}}e_{s_1\cdots s_i\cdots s_n} \nonumber\\
&\sum_{l=\sum_{j=1}^n s_j+1}^{\sum_{j=1}^n g_j+1} {\sum_{j=1}^n g_j+1 \choose l}y^{l}(1-y)^{\sum_{j=1}^n g_j+1-l}dy \nonumber \\
=&\frac{1}{(\sum_{j=1}^n g_j+1)(\sum_{j=1}^n g_j+2)}\sum_{s_1=0}^{g_1}\cdots \sum_{s_i=0}^{g_i}\cdots \sum_{s_n=0}^{g_n} \frac{\prod_{j=1}^n {g_j \choose s_j}}{{\sum_{j=1}^n g_j \choose \sum_{j=1}^n s_j}}e_{s_1\cdots s_i\cdots s_n} \nonumber\\
&\sum_{l=\sum_{j=1}^n s_j+1}^{\sum_{j=1}^n g_j+1}\sum_{m=l+1}^{\sum_{j=1}^n g_j+2} {\sum_{j=1}^n g_j+2 \choose m}x^{m}(1-x)^{\sum_{j=1}^n g_j+2-m} \nonumber \\
=&\frac{1}{(k+1)(k+2)}\sum_{m=0}^{k+2}{k+2 \choose m}x^m(1-x)^{k+2-m} \nonumber \\
&\sum_{l=0}^{m-1}\sum_{s_1=0}^{l-1}\cdots\sum_{s_i=0}^{l-1-\sum_{j=1}^{i-1}s_j}\cdots\sum_{s_n=0}^{l-1-\sum_{j=1}^{n-1}s_j}
\frac{\prod_{j=1}^n {g_j \choose s_j}}{{k \choose \sum_{j=1}^n s_j}}e_{s_1\cdots s_i\cdots s_n}
\end{align}
and
\begin{align}
\int_0^1\int_0^yH(r)drdy
=&\frac{1}{(k+1)(k+2)}\sum_{l=0}^{k+1}\sum_{s_1=0}^{l-1}\cdots\sum_{s_i=0}^{l-1-\sum_{j=1}^{i-1}s_j}\cdots\sum_{s_n=0}^{l-1-\sum_{j=1}^{n-1}s_j}
\frac{\prod_{j=1}^n {g_j \choose s_j}}{{k \choose \sum_{j=1}^n s_j}}e_{s_1\cdots s_i\cdots s_n} \nonumber \\
=&\frac{1}{(k+1)(k+2)}\sum_{l=0}^{k}\sum_{s_1=0}^{l}\cdots\sum_{s_i=0}^{l-\sum_{j=1}^{i-1}s_j}\cdots\sum_{s_n=0}^{l-\sum_{j=1}^{n-1}s_j}
\frac{\prod_{j=1}^n {g_j \choose s_j}}{{k \choose \sum_{j=1}^n s_j}}e_{s_1\cdots s_i\cdots s_n} \nonumber \\
=&\frac{1}{(k+1)(k+2)}\sum_{s_1=0}^{g_1}\cdots\sum_{s_i=0}^{g_i}\cdots\sum_{s_n=0}^{g_n}
\frac{\prod_{j=1}^n {g_j \choose s_j}}{{k \choose \sum_{j=1}^n s_j}}\left(k+1-\sum_{j=1}^n s_j\right)e_{s_1\cdots s_i\cdots s_n} \nonumber.
\end{align}

Extending Eq~(\ref{interal1}) and taking $x=1/N$ ($N\gg 1$), we get
\begin{align}
&\frac{1}{(k+1)(k+2)}\left[0+0+{k+2 \choose 2}x^2(1-x)^{k}e_{0\cdots 0\cdots 0}+\cdots\right] \nonumber \\
=&\frac{1}{(k+1)(k+2)}{k+2 \choose 2}x^2(1-x)^{k}e_{0\cdots 0\cdots 0}+O(x^3) \nonumber \\
=&\frac{e_{0\cdots 0\cdots 0}}{2N^2}+O(\frac{1}{N^3})
\end{align}

Finally, we get the fixation probability $\rho_A=\phi_A(1/N)$ for $N\gg 1$, given by
\begin{align} \label{fixation_probabilityA}
\rho_A\approx
&\frac{1}{N}+\frac{\omega N(k-1)}{k^2}\Big[
\frac{1}{N(k+1)(k+2)}\sum_{s_1=0}^{g_1}\cdots\sum_{s_i=0}^{g_i}\cdots\sum_{s_n=0}^{g_n}
\frac{\prod_{j=1}^n {g_j \choose s_j}}{{k \choose \sum_{j=1}^n s_j}}\left(k+1-\sum_{j=1}^n s_j\right)e_{s_1\cdots s_i\cdots s_n}
\nonumber \\
&-\frac{e_{0\cdots 0\cdots 0}}{2N^2}\Big] \nonumber\\
\approx &\frac{1}{N}+\frac{\omega (k-1)}{k^2(k+1)(k+2)}\sum_{s_1=0}^{g_1}\cdots\sum_{s_i=0}^{g_i}\cdots\sum_{s_n=0}^{g_n}
\frac{\prod_{j=1}^n {g_j \choose s_j}}{{k \choose \sum_{j=1}^n s_j}}\left(k+1-\sum_{j=1}^n s_j\right)e_{s_1\cdots s_i\cdots s_n}
\end{align}

\subsection{Fixation probabilities, sigma rule and structure coefficients}
Equation~(\ref{fixation_probabilityA}) shows that $\rho_A>\frac{1}{N}$ only if
\begin{align} \label{simple_cariteria}
\sum_{s_1=0}^{g_1}\cdots\sum_{s_i=0}^{g_i}\cdots\sum_{s_n=0}^{g_n}\frac{\prod_{j=1}^n {g_j \choose s_j}}{{k \choose \sum_{j=1}^n s_j}}\left(k+1-\sum_{j=1}^n s_j\right)e_{s_1\cdots s_i\cdots s_n}>0
\end{align}
From Eqs~(\ref{eji}-\ref{dji}), $e_{s_1\cdots s_i\cdots s_n}$ is linear in $a_{s_1\cdots s_i\cdots s_n}$ and $b_{s_1\cdots s_i\cdots s_n}$.
Thus the left side of formula (\ref{simple_cariteria}) is linear in $a_{s_1\cdots s_i\cdots s_n}$ and $b_{s_1\cdots s_i\cdots s_n}$, implying that there are $\alpha_{s_1\cdots s_i\cdots s_n}$
and $\beta_{s_1\cdots s_i\cdots s_n}$ such that
\begin{align} \label{alpha_beta}
&\sum_{s_1=0}^{g_1}\cdots\sum_{s_i=0}^{g_i}\cdots\sum_{s_n=0}^{g_n}\frac{\prod_{j=1}^n {g_j \choose s_j}}{{k \choose \sum_{j=1}^n s_j}}\left(k+1-\sum_{j=1}^n s_j\right)e_{s_1\cdots s_i\cdots s_n} \nonumber \\
=&\sum_{s_1=0}^{g_1}\cdots\sum_{s_i=0}^{g_i}\cdots\sum_{s_n=0}^{g_n}\left(\alpha_{s_1\cdots s_i\cdots s_n}a_{s_1\cdots s_i\cdots s_n}+\beta_{s_1\cdots s_i\cdots s_n}b_{s_1\cdots s_i\cdots s_n}\right)
\end{align}
Hence, we rewrite $\rho_A$ and the fixation probability of a single mutant $\rho_B$ as
\begin{align}
\rho_A \approx
&\frac{1}{N}+\frac{\omega (k-1)}{k^2(k+1)(k+2)} \nonumber\\
&\sum_{s_1=0}^{g_1}\cdots\sum_{s_i=0}^{g_i}\cdots\sum_{s_n=0}^{g_n}\left(\alpha_{s_1\cdots s_i\cdots s_n}a_{s_1\cdots s_i\cdots s_n}+\beta_{s_1\cdots s_i\cdots s_n}b_{s_1\cdots s_i\cdots s_n}\right)  \label{rhoA}
\end{align}
\begin{align}
\rho_B \approx
&\frac{1}{N}+\frac{\omega (k-1)}{k^2(k+1)(k+2)}
\sum_{s_1=0}^{g_1}\cdots\sum_{s_i=0}^{g_i}\cdots\sum_{s_n=0}^{g_n}\Big(\alpha_{s_1\cdots s_i\cdots s_n}b_{(g_1-s_1)\cdots (g_i-s_i)\cdots (g_n-s_n)} \nonumber \\
& \hspace{7cm} +\beta_{s_1\cdots s_i\cdots s_n}a_{(g_1-s_1)\cdots (g_i-s_i)\cdots (g_n-s_n)}\Big)  \label{rhoB}
\end{align}
Thus, under weak selection, we have
\begin{align}
& \rho_A>\rho_B \nonumber \\
\Longleftrightarrow
& \sum_{s_1=0}^{g_1}\cdots\sum_{s_i=0}^{g_i}\cdots\sum_{s_n=0}^{g_n}\left(\alpha_{s_1\cdots s_i\cdots s_n}a_{s_1\cdots s_i\cdots s_n}+\beta_{s_1\cdots s_i\cdots s_n}b_{s_1\cdots s_i\cdots s_n}\right) \nonumber\\
& >
\sum_{s_1=0}^{g_1}\cdots\sum_{s_i=0}^{g_i}\cdots\sum_{s_n=0}^{g_n}\left(\alpha_{s_1\cdots s_i\cdots s_n}b_{(g_1-s_1)\cdots (g_i-s_i)\cdots (g_n-s_n)}+\beta_{s_1\cdots s_i\cdots s_n}a_{(g_1-s_1)\cdots (g_i-s_i)\cdots (g_n-s_n)}\right)  \nonumber \\
\Longleftrightarrow
& \sum_{s_1=0}^{g_1}\cdots\sum_{s_i=0}^{g_i}\cdots\sum_{s_n=0}^{g_n}\left(\alpha_{s_1\cdots s_i\cdots s_n}-
\beta_{(g_1-s_1)\cdots (g_i-s_i)\cdots (g_n-s_n)}\right)a_{s_1\cdots s_i\cdots s_n} \nonumber\\
&+\left(\beta_{(g_1-s_1)\cdots (g_i-s_i)\cdots (g_n-s_n)}-\alpha_{s_1\cdots s_i\cdots s_n}\right)b_{(g_1-s_1)\cdots (g_i-s_i)\cdots (g_n-s_n)}>0 \nonumber\\
\Longleftrightarrow
&\sum_{s_1=0}^{g_1}\cdots\sum_{s_i=0}^{g_i}\cdots\sum_{s_n=0}^{g_n}\left(\alpha_{s_1\cdots s_i\cdots s_n}-
\beta_{(g_1-s_1)\cdots (g_i-s_i)\cdots (g_n-s_n)}\right)\left(a_{s_1\cdots s_i\cdots s_n}-b_{(g_1-s_1)\cdots (g_i-s_i)\cdots (g_n-s_n)}\right)>0 \nonumber \\
\Longleftrightarrow
&\sum_{s_1=0}^{g_1}\cdots\sum_{s_i=0}^{g_i}\cdots\sum_{s_n=0}^{g_n}\sigma_{s_1\cdots s_i\cdots s_n}\left(a_{s_1\cdots s_i\cdots s_n}-b_{(g_1-s_1)\cdots (g_i-s_i)\cdots (g_n-s_n)}\right)>0 \label{sigma_rule}
\end{align}
Equation~(\ref{sigma_rule}) is termed "sigma rule" and its coefficients
\begin{align}
\sigma_{s_1\cdots s_i\cdots s_n}=\alpha_{s_1\cdots s_i\cdots s_n}-
\beta_{(g_1-s_1)\cdots (g_i-s_i)\cdots (g_n-s_n)} \label{sigma}
\end{align}
are the structure coefficients.
Here we refer to the method in Ref.~\cite{2016-Pena-p1-15} to calculate $\alpha_{s_1\cdots s_i\cdots s_n}$ and $\beta_{s_1\cdots s_i\cdots s_n}$.
For a multi-player game with $a_{s_1\cdots s_i\cdots s_n}=\prod_{i=1}^n\delta_{\tilde{s}_i,s_i}$ (only $a_{\tilde{s}_1\cdots \tilde{s}_i\cdots \tilde{s}_n}=1$ and all others are $0$) and $b_{s_1\cdots s_i\cdots s_n}=0$, by Eq~(\ref{alpha_beta}), we have
\begin{align} \label{alpha1}
\alpha_{\tilde{s}_1\cdots \tilde{s}_i\cdots \tilde{s}_n}
=&\sum_{s_1=0}^{g_1}\cdots\sum_{s_i=0}^{g_i}\cdots\sum_{s_n=0}^{g_n}\frac{\prod_{j=1}^n {g_j \choose s_j}}{{k \choose \sum_{j=1}^n s_j}}\left(k+1-\sum_{j=1}^n s_j\right)e_{s_1\cdots s_i\cdots s_n}^{\tilde{s}_1\cdots \tilde{s}_i\cdots \tilde{s}_n},
\end{align}
where $e_{s_1\cdots s_i\cdots s_n}^{\tilde{s}_1\cdots \tilde{s}_i\cdots \tilde{s}_n}$ denotes the coefficient $e_{s_1\cdots s_i\cdots s_n}$ with $a_{s_1\cdots s_i\cdots s_n}=\prod_{i=1}^n\delta_{\tilde{s}_i,s_i}$ and $b_{s_1\cdots s_i\cdots s_n}=0$ for any combination $s_1\cdots s_i\cdots s_n$.
Analogously, $\leftidx{^i}c_{s_1\cdots s_i\cdots s_n}^{\tilde{s}_1\cdots\tilde{s}_i\cdots\tilde{s}_n}$ and $\leftidx{^i}d_{s_1\cdots s_i\cdots s_n}^{\tilde{s}_1\cdots\tilde{s}_i\cdots\tilde{s}_n}$ respectively correspond to $\leftidx{^i}c_{s_1\cdots s_i\cdots s_n}$ and $\leftidx{^i}d_{s_1\cdots s_i\cdots s_n}$ with $a_{s_1\cdots s_i\cdots s_n}=\prod_{i=1}^n\delta_{\tilde{s}_i,s_i}$ and $b_{s_1\cdots s_i\cdots s_n}=0$.
From Eqs~(\ref{cji}) and (\ref{dji}), we have
\begin{align}
\leftidx{^i}c_{s_1\cdots s_i\cdots s_n}^{\tilde{s}_1\cdots\tilde{s}_i\cdots\tilde{s}_n}
=&\prod_{j=1}^n {g_i-\delta_{i,j}-s_i \choose \tilde{s}_i-s_i}z^{\tilde{s}_i-s_i}(1-z)^{g_i-\delta_{i,j}-\tilde{s}_i} \nonumber\\
=&{g_1-s_1 \choose g_1-\tilde{s}_1}\cdots{g_i-1-s_i \choose g_i-1-\tilde{s}_i}\cdots{g_n-s_n \choose g_n-\tilde{s}_n}
\frac{(k-2)^{k-1-\sum_{j=1}^n \tilde{s}_j}}{(k-1)^{k-1-\sum_{j=1}^n s_j}}, \label{cji2}
\end{align}
%%%%%
\begin{align}
\leftidx{^i}d_{s_1\cdots s_i\cdots s_n}^{\tilde{s}_1\cdots\tilde{s}_i\cdots\tilde{s}_n}
=&\prod_{j=1}^n {g_i-\delta_{i,j}-s_i \choose \tilde{s}_i-\delta_{i,j}-s_i}z^{\tilde{s}_i-\delta_{i,j}-s_i}(1-z)^{g_i-\tilde{s}_i} \nonumber\\
=&{g_1-s_1 \choose g_1-\tilde{s}_1}\cdots{g_i-1-s_i \choose g_i-\tilde{s}_i}\cdots{g_n-s_n \choose g_n-\tilde{s}_n}
\frac{(k-2)^{k-\sum_{j=1}^n \tilde{s}_j}}{(k-1)^{k-1-\sum_{j=1}^n s_j}}. \label{dji2}
\end{align}
Substituting Eqs~(\ref{eji},\ref{cji2},\ref{dji2}) into Eq~(\ref{alpha1}), we obtain
\begin{align}
\alpha_{\tilde{s}_1\cdots \tilde{s}_i\cdots \tilde{s}_n}
=&\sum_{s_1=0}^{g_1}\cdots\sum_{s_i=0}^{g_i}\cdots\sum_{s_n=0}^{g_n}\frac{\prod_{j=1}^n {g_j \choose s_j}}{{k \choose \sum_{j=1}^n s_j}}\left(k+1-\sum_{j=1}^n s_j\right) \nonumber\\
&\sum_{i=1}^n \left[zs_i\leftidx{^i}c_{s_1\cdots (s_i-1)\cdots s_n}^{\tilde{s}_1\cdots\tilde{s}_i\cdots\tilde{s}_n}+(g_i-s_i)\leftidx{^i}c_{s_1\cdots s_i\cdots s_n}^{\tilde{s}_1\cdots\tilde{s}_i\cdots\tilde{s}_n}+s_i\leftidx{^i}d_{s_1\cdots (s_i-1)\cdots s_n}^{\tilde{s}_1\cdots\tilde{s}_i\cdots\tilde{s}_n}+z(g_i-s_i)\leftidx{^i}d_{s_1\cdots s_i\cdots s_n}^{\tilde{s}_1\cdots\tilde{s}_i\cdots\tilde{s}_n}\right] \nonumber\\
=&\sum_{s_1=0}^{g_1}\cdots\sum_{s_i=0}^{g_i}\cdots\sum_{s_n=0}^{g_n}\sum_{i=1}^n
\Bigg[\frac{\prod_{j=1}^n {g_j \choose s_j+\delta_{i,j}}}{{k \choose \sum_{j=1}^n (s_j+\delta_{i,j})}}
\left(k-\sum_{j=1}^n s_j\right)(s_i+1)z\leftidx{^i}c_{s_1\cdots s_i\cdots s_n}^{\tilde{s}_1\cdots\tilde{s}_i\cdots\tilde{s}_n} \nonumber\\
&\hspace{4cm} +\frac{\prod_{j=1}^n {g_j \choose s_j}}{{k \choose \sum_{j=1}^n s_j}}\left(k+1-\sum_{j=1}^n s_j\right)(g_i-s_i)\leftidx{^i}c_{s_1\cdots s_i\cdots s_n}^{\tilde{s}_1\cdots\tilde{s}_i\cdots\tilde{s}_n}
\nonumber\\
&\hspace{4cm} +\frac{\prod_{j=1}^n {g_j \choose s_j+\delta_{i,j}}}{{k \choose \sum_{j=1}^n (s_j+\delta_{i,j})}}
\left(k-\sum_{j=1}^n s_j\right)(s_i+1)\leftidx{^i}d_{s_1\cdots s_i\cdots s_n}^{\tilde{s}_1\cdots\tilde{s}_i\cdots\tilde{s}_n} \nonumber\\
&\hspace{4cm} +\frac{\prod_{j=1}^n {g_j \choose s_j}}{{k \choose \sum_{j=1}^n s_j}}\left(k+1-\sum_{j=1}^ns_j\right)
(g_i-s_i)z\leftidx{^i}d_{s_1\cdots s_i\cdots s_n}^{\tilde{s}_1\cdots\tilde{s}_i\cdots\tilde{s}_n}\Bigg] \nonumber \\
=&\sum_{s_1=0}^{g_1}\cdots\sum_{s_i=0}^{g_i}\cdots\sum_{s_n=0}^{g_n}\frac{\prod_{j=1}^n {g_j \choose s_j}}{{k \choose \sum_{j=1}^n s_j}}\sum_{i=1}^n\Bigg[\frac{g_i-s_i}{k-1}\left(k^2-(k-2)\sum_{j=1}^ns_j\right)\leftidx{^i}c_{s_1\cdots s_i\cdots s_n}^{\tilde{s}_1\cdots\tilde{s}_i\cdots\tilde{s}_n} \nonumber\\
&\hspace{5.5cm} +\frac{g_i-s_i}{k-1}\left(2k+(k-2)\sum_{j=1}^ns_j\right)\leftidx{^i}d_{s_1\cdots s_i\cdots s_n}^{\tilde{s}_1\cdots\tilde{s}_i\cdots\tilde{s}_n} \Bigg] \label{alpha_beta2}\\
=&\frac{(k-2)^{k-1-\sum_{j=1}^n\tilde{s}_j}}{k-1}\sum_{s_1=0}^{g_1}\cdots\sum_{s_i=0}^{g_i}\cdots\sum_{s_n=0}^{g_n}\frac{\prod_{j=1}^n {g_j \choose s_j}{g_j-s_j \choose g_j-\tilde{s}_j}}{{k \choose \sum_{j=1}^n s_j}} \nonumber \\
&\Bigg[\frac{\left(k^2-(k-2)\sum_{j=1}^ns_j\right)\left(k-\sum_{j=1}^n\tilde{s}_j\right)}{(k-1)^{k-1-\sum_{j=1}^ns_j}} \nonumber \\
&+\frac{(k-2)\left(2k+(k-2)\sum_{j=1}^ns_j\right)\left(\sum_{j=1}^n\tilde{s}_j-\sum_{j=1}^ns_j\right)}{(k-1)^{k-1-\sum_{j=1}^ns_j}}\Bigg] \label{alpha2}
%\nonumber
\end{align}

We make full use of a following identity to simply Eq~(\ref{alpha2}):
\begin{align} \label{identity_sigman}
&\sum_{s_1=0}^{g_1}\sum_{s_2=0}^{g_2}\cdots\sum_{s_n=0}^{g_n}\frac{\prod_{j=1}^n {g_j \choose s_j}{g_j-s_j \choose g_j-\tilde{s}_j}}{{\sum_{j=1}^n g_j \choose \sum_{j=1}^n s_j}}\Theta\left(\sum_{j=1}^ns_j,\sum_{j=1}^n\tilde{s}_j\right) \nonumber \\
=&\frac{\prod_{j=1}^n{g_j \choose \tilde{s}_j}}{{\sum_{j=1}^n g_j \choose \sum_{j=1}^n\tilde{s}_j}}\sum_{l=0}^{\sum_{j=1}^n g_j}{\sum_{j=1}^n g_j-l \choose \sum_{j=1}^n g_j-\sum_{j=1}^n\tilde{s}_j}\Theta\left(l,\sum_{j=1}^n\tilde{s}_j\right),
\end{align}
where $\Theta\left(\sum_{j=1}^ns_j,\sum_{j=1}^n\tilde{s}_j\right)$ is a function of $\sum_{j=1}^ns_j$ and $\sum_{j=1}^n\tilde{s}_j$.
We here give a brief proof for this identity.
First we investigate the case with $n=2$.
\begin{align}
&\sum_{s_1=0}^{g_1}\sum_{s_2=0}^{g_2}\frac{{g_1 \choose s_1}{g_1-s_1 \choose g_1-\tilde{s}_1}{g_2 \choose s_2}{g_2-s_2 \choose g_2-\tilde{s}_2}}{{g_1+g_2 \choose s_1+s_2}}\Theta(s_1+s_2,\tilde{s}_1+\tilde{s}_2) \nonumber\\
=&\sum_{s_1=0}^{g_1}\sum_{s_2=0}^{g_2}\frac{g_1!}{s_1!(g_1-s_1)!}\frac{g_2!}{s_2!(g_2-s_2)!}\frac{(g_1-s_1)!}{(g_1-\tilde{s}_1)!(\tilde{s}_1-s_1)!}
\frac{(g_2-s_2)!}{(g_2-\tilde{s}_2)!(\tilde{s}_2-s_2)!} \nonumber \\
&\qquad \quad\frac{(s_1+s_2)!(g_1+g_2-s_1-s_2)!}{(g_1+g_2)!}\Theta(s_1+s_2,\tilde{s}_1+\tilde{s}_2) \nonumber\\
=&\sum_{s_1=0}^{g_1}\sum_{s_2=0}^{g_2}\frac{g_1!g_2!}{(g_1+g_2)!}\frac{(s_1+s_2)!}{s_1!s_2!}\frac{(\tilde{s}_1+\tilde{s}_2-s_1-s_2)!}{(\tilde{s}_1-s_1)!(\tilde{s}_2-s_2)!}
\frac{(g_1+g_2-\tilde{s}_1-\tilde{s}_2)!}{(g_1-\tilde{s}_1)!(g_2-\tilde{s}_2)!}\nonumber \\
&\qquad \quad\frac{(g_1+g_2-s_1-s_2)!}{(g_1+g_2-\tilde{s}_1-\tilde{s}_2)!(\tilde{s}_1+\tilde{s}_2-s_1-s_2)!}\Theta(s_1+s_2,\tilde{s}_1+\tilde{s}_2) \nonumber\\
=&\frac{1}{{g_1+g_2 \choose g_1}}\sum_{s_1=0}^{g_1}\sum_{s_2=0}^{g_2}{s_1+s_2 \choose s_1}{\tilde{s}_1+\tilde{s}_2-s_1-s_2 \choose \tilde{s}_1-s_1}{g_1+g_2-\tilde{s}_1-\tilde{s}_2 \choose g_1-\tilde{s}_1}{g_1+g_2-s_1-s_2 \choose g_1+g_2-\tilde{s}_1-\tilde{s}_2}\nonumber\\
&\hspace{3cm}\Theta(s_1+s_2,\tilde{s}_1+\tilde{s}_2) \nonumber \\
=&\frac{1}{{g_1+g_2 \choose g_1}}\sum_{l=0}^{g_1+g_2}\sum_{s_1=0}^{l}{l \choose s_1}{\tilde{s}_1+\tilde{s}_2-l \choose \tilde{s}_1-s_1}{g_1+g_2-\tilde{s}_1-\tilde{s}_2 \choose g_1-\tilde{s}_1}{g_1+g_2-l \choose g_1+g_2-\tilde{s}_1-\tilde{s}_2}\Theta(l,\tilde{s}_1+\tilde{s}_2) \nonumber \\
=&\frac{1}{{g_1+g_2 \choose g_1}}\sum_{l=0}^{g_1+g_2}{\tilde{s}_1+\tilde{s}_2 \choose \tilde{s}_1}{g_1+g_2-\tilde{s}_1-\tilde{s}_2 \choose g_1-\tilde{s}_1}{g_1+g_2-l \choose g_1+g_2-\tilde{s}_1-\tilde{s}_2}\Theta(l,\tilde{s}_1+\tilde{s}_2) \nonumber \\
=&\frac{{g_1 \choose \tilde{s}_1}{g_2 \choose \tilde{s}_2}}{{g_1+g_2 \choose \tilde{s}_1+\tilde{s}_2}}\sum_{l=0}^{g_1+g_2}{g_1+g_2-l \choose g_1+g_2-\tilde{s}_1-\tilde{s}_2}\Theta(l,\tilde{s}_1+\tilde{s}_2). \label{identity_sigma}
\end{align}
Then we extend the identity in case $n=2$ to any $n$.
We decompose this long equation (see terms in square brackets) and use Eq~(\ref{identity_sigma}) repeatedly.
Finally, we can complete the proof and obtain Eq~(\ref{identity_sigman}).
\begin{align}
&\sum_{s_1=0}^{g_1}\sum_{s_2=0}^{g_2}\cdots \sum_{s_n=0}^{g_n}\frac{{g_1 \choose s_1}{g_2 \choose s_2}\cdots {g_n \choose s_n}}
{{g_1+g_2+ \cdots +g_n \choose s_1+s_2+\cdots +s_n}}{g_1-s_1 \choose g_1-\tilde{s}_1}{g_2-s_2 \choose g_2-\tilde{s}_2}\cdots{g_n-s_n \choose g_n-\tilde{s}_n}
\Theta\left(\sum_{j=1}^n s_j,\sum_{j=1}^n \tilde{s}_{j}\right) \nonumber\\
=&\sum_{s_1=0}^{g_1}\sum_{s_2=0}^{g_2}\cdots \sum_{s_{n-2}=0}^{g_{n-2}}\left[\sum_{s_{n-1}=0}^{g_{n-1}}\sum_{s_n=0}^{g_n}
\frac{{g_{n-1} \choose s_{n-1}}{g_{n} \choose s_n}}{{g_{n-1}+g_n \choose s_{n-1}+s_n}}{g_{n-1}-s_{n-1} \choose g_{n-1}-\tilde{s}_{n-1}}
{g_n-s_n \choose g_n-\tilde{s}_n}\right]\frac{{g_1 \choose s_1}{g_2 \choose s_2}\cdots {g_{n-2} \choose s_{n-2}}}{{g_1+g_2+ \cdots +g_n \choose s_1+s_2+\cdots +s_{n-2}+(s_{n-1}+s_n)}} \nonumber\\
&\cdot{g_{n-1}+g_n \choose s_{n-1}+s_n}{g_1-s_1 \choose g_1-\tilde{s}_1}{g_2-s_2 \choose g_2-\tilde{s}_2}\cdots{g_{n-2}-s_{n-2} \choose g_{n-2}-\tilde{s}_{n-2}}
\Theta\left(\sum_{j=0}^{n-2} s_j+(s_{n-1}+s_n),\sum_{j=1}^n \tilde{s}_{j}\right) \nonumber \\
=&\sum_{s_1=0}^{g_1}\sum_{s_2=0}^{g_2}\cdots \sum_{s_{n-2}=0}^{g_{n-2}}
\left[\frac{{g_{n-1} \choose \tilde{s}_{n-1}}{g_{n} \choose \tilde{s}_n}}{{g_{n-1}+g_n \choose \tilde{s}_{n-1}+\tilde{s}_n}}\sum_{l=0}^{g_{n-1}+g_{n}}
{g_{n-1}+g_{n}-l \choose {g_{n-1}+g_{n}-\tilde{s}_{n-1}-\tilde{s}_n}}\right]
\frac{{g_1 \choose s_1}{g_2 \choose s_2}\cdots {g_{n-2} \choose s_{n-2}}}{{g_1+g_2+ \cdots +g_n \choose s_1+s_2+\cdots +s_{n-2}+l}} \nonumber\\
&\cdot{g_{n-1}+g_n \choose l}{g_1-s_1 \choose g_1-\tilde{s}_1}{g_2-s_2 \choose g_2-\tilde{s}_2}\cdots{g_{n-2}-s_{n-2} \choose g_{n-2}-\tilde{s}_{n-2}}
\Theta\left(\sum_{j=0}^{n-2} s_j+l,\sum_{j=1}^n \tilde{s}_{j}\right) \nonumber \\
=&\frac{{g_{n-1} \choose \tilde{s}_{n-1}}{g_{n} \choose \tilde{s}_n}}{{g_{n-1}+g_n \choose \tilde{s}_{n-1}+\tilde{s}_n}}
\sum_{s_1=0}^{g_1}\sum_{s_2=0}^{g_2}\cdots \sum_{s_{n-3}=0}^{g_{n-3}}\Bigg[\sum_{s_{n-2}=0}^{g_{n-2}}\sum_{l=0}^{g_{n-1}+g_{n}}
\frac{{g_{n-2} \choose s_{n-2}}{g_{n-1}+g_n \choose l}}{{g_{n-2}+g_{n-1}+g_n \choose l+s_{n-2}}}
{g_{n-1}+g_{n}-l \choose {g_{n-1}+g_{n}-\tilde{s}_{n-1}-\tilde{s}_n}}\nonumber\\
&{g_{n-2}-s_{n-2} \choose g_{n-2}-\tilde{s}_{n-2}}\Bigg]
\cdot{g_{n-2}+g_{n-1}+g_n \choose l+s_{n-2}}\frac{{g_1 \choose s_1}{g_2 \choose s_2}\cdots {g_{n-3} \choose s_{n-3}}}{{g_1+g_2+ \cdots +g_n \choose s_1+s_2+\cdots +(s_{n-2}+l)}}
{g_1-s_1 \choose g_1-\tilde{s}_1}{g_2-s_2 \choose g_2-\tilde{s}_2}\cdots{g_{n-3}-s_{n-3} \choose g_{n-3}-\tilde{s}_{n-3}} \nonumber \\
&\cdot\Theta\left(\sum_{j=0}^{n-3} s_j+(s_{n-2}+l),\sum_{j=1}^n \tilde{s}_{j}\right) \nonumber \\
=& \cdots\cdots \nonumber\\
=&\frac{\prod_{j=1}^n{g_j \choose \tilde{s}_j}}{{\sum_{j=1}^n g_j \choose \sum_{j=1}^n\tilde{s}_j}}\sum_{l=0}^{\sum_{j=1}^n g_j}{\sum_{j=1}^n g_j-l \choose \sum_{j=1}^n g_j-\sum_{j=1}^n\tilde{s}_j}\Theta\left(l,\sum_{j=1}^n\tilde{s}_j\right) \nonumber
\end{align}

Applying Eq~(\ref{identity_sigman}) to Eq~(\ref{alpha2}) and taking $\sum_{j=1}^n g_j=k$, we have
\begin{align}
\alpha_{\tilde{s}_1\tilde{s}_2\cdots \tilde{s}_n}
=&\frac{(k-2)^{k-1-\sum_{j=1}^{n}\tilde{s}_j}}
{k-1}\frac{\Pi_{j=1}^n{g_j \choose \tilde{s}_j}}
{{k \choose \sum_{j=1}^n\tilde{s}_j}}
\sum_{l=0}^{k}(k-l)\Bigg[{k-1-l \choose k-1-\sum_{j=1}^n\tilde{s}_j}\frac{k^2-(k-2)l}{(k-1)^{k-1-l}} \nonumber \\
&+{k-1-l \choose k-\sum_{j=1}^n\tilde{s}_j}\frac{(2k+(k-2)l)(k-2)}{(k-1)^{k-1-l}}\Bigg]. \label{alpha_final}
\end{align}

Analogously, we use a multi-player game with $a_{s_1\cdots s_i\cdots s_n}=0$ and $b_{s_1\cdots s_i\cdots s_n}=\prod_{i=1}^n\delta_{\tilde{s}_i,s_i}$ (only $b_{\tilde{s}_1\cdots \tilde{s}_i\cdots \tilde{s}_n}=1$ and all others are $0$) to calculate $\beta_{\tilde{s}_1\cdots \tilde{s}_i\cdots \tilde{s}_n}$.
$\beta_{\tilde{s}_1\cdots \tilde{s}_i\cdots \tilde{s}_n}$ also has the form of Eq~(\ref{alpha_beta2}).
Using Eq~(\ref{alpha_beta}) and referring to Eqs~(\ref{alpha1}-\ref{alpha2}), we have
\begin{align}
\leftidx{^i}c_{s_1\cdots s_i\cdots s_n}^{\tilde{s}_1\cdots\tilde{s}_i\cdots\tilde{s}_n}
=&-\prod_{j=1}^n {s_j \choose s_j-\tilde{s}_j}z^{s_j-\tilde{s}_j}(1-z)^{\tilde{s}_j} \nonumber\\
=&-{s_1 \choose s_1-\tilde{s}_1}\cdots{s_i \choose s_i-\tilde{s}_i}\cdots{s_n \choose s_n-\tilde{s}_n}
\frac{(k-2)^{\sum_{j=1}^n \tilde{s}_j}}{(k-1)^{\sum_{j=1}^n s_j}}, \label{cji3}
\end{align}
%%%%%
\begin{align}
\leftidx{^i}d_{s_1\cdots s_i\cdots s_n}^{\tilde{s}_1\cdots\tilde{s}_i\cdots\tilde{s}_n}
=&-\prod_{j=1}^n {s_j \choose s_j-\tilde{s}_j+\delta_{i,j}}z^{s_j-\tilde{s}_j+\delta_{i,j}}(1-z)^{\tilde{s}_j-\delta_{i,j}} \nonumber\\
=&-{s_1 \choose s_1-\tilde{s}_1}\cdots{s_i \choose s_i-\tilde{s}_i+1}\cdots{s_n \choose s_n-\tilde{s}_n}
\frac{(k-2)^{\sum_{j=1}^n \tilde{s}_j-1}}{(k-1)^{\sum_{j=1}^n s_j}}. \label{dji3}
\end{align}

Here we give two identities, which can be derived in an analogous way to Eq~(\ref{identity_sigman}), i.e.,
\begin{align}
&\sum_{s_1=0}^{g_1}\sum_{s_2=0}^{g_2}\cdots\sum_{s_n=0}^{g_n}\frac{\prod_{j=1}^n {g_j \choose s_j}{s_j \choose s_j-\tilde{s}_j}}{{\sum_{j=1}^n g_j \choose \sum_{j=1}^n s_j}}\Theta\left(\sum_{j=1}^ns_j,\sum_{j=1}^n\tilde{s}_j\right) \nonumber \\
=&\frac{\prod_{j=1}^n{g_j \choose \tilde{s}_j}}{{\sum_{j=1}^n g_j \choose \sum_{j=1}^n\tilde{s}_j}}\sum_{l=0}^{\sum_{j=1}^n g_j}{l \choose \sum_{j=1}^n\tilde{s}_j}\Theta\left(l,\sum_{j=1}^n\tilde{s}_j\right) \label{identity_sigman2}
\end{align}
and
\begin{align}
&\sum_{s_1=0}^{g_1}\sum_{s_2=0}^{g_2}\cdots\sum_{s_n=0}^{g_n}\frac{\prod_{j=1}^n {g_j \choose s_j}{s_j \choose s_j-\tilde{s}_j+\delta_{i,j}}}{{\sum_{j=1}^n g_j \choose \sum_{j=1}^n s_j}}(g_i-s_i)\Theta\left(\sum_{j=1}^ns_j,\sum_{j=1}^n\tilde{s}_j\right) \nonumber \\
=&\frac{\tilde{s}_i}{\sum_{j}\tilde{s}_j}\frac{\prod_{j=1}^n{g_j \choose \tilde{s}_j}}{{\sum_{j=1}^n g_j \choose \sum_{j=1}^n\tilde{s}_j}}\sum_{l=0}^{\sum_{j=1}^n g_j}{l \choose \sum_{j=1}^n\tilde{s}_j-1}\Theta\left(l,\sum_{j=1}^n\tilde{s}_j\right) \label{identity_sigman3}
\end{align}

Substituting Eqs~(\ref{cji3}) and (\ref{dji3}) into Eq~(\ref{alpha_beta2}) and applying Eqs~(\ref{identity_sigman2},\ref{identity_sigman3}), we have
\begin{align}
\beta_{\tilde{s}_1\tilde{s}_2\cdots \tilde{s}_n}=
&-\frac{(k-2)^{\sum_{j=1}^n\tilde{s}_j}}{k-1}\frac{\Pi_{j=1}^n{g_j \choose \tilde{s}_j}}
{{k \choose \sum_{j=1}^n\tilde{s}_j}}
\sum_{l=0}^{k}(k-l)\Bigg[{l \choose \sum_{j=1}^n\tilde{s}_j}\frac{k^2-(k-2)l}{(k-1)^l}\nonumber\\
&+{l \choose \sum_{j=1}^n\tilde{s}_j-1}\frac{2k+(k-2)l}{(k-2)(k-1)^l} \Bigg] \label{beta2}
\end{align}
Substituting Eq~(\ref{alpha_final}) and Eq~(\ref{beta2}) into Eq~(\ref{sigma}), we have
\begin{align}
\sigma_{s_1s_2\cdots s_n}=
&\frac{(k-2)^{(k-\sum_{j=1}^ns_j)}}{k-1}\frac{\Pi_{j=1}^n{g_j \choose s_j}}
{{k \choose \sum_{j=1}^ns_j}}\nonumber \\
&\sum_{l=0}^{k}(k-l)\left\{\left[k^2-(k-2)l\right]\Phi\left(k,\sum_{j=1}^ns_j,l\right)+\left[2k+(k-2)l\right]\Psi\left(k,\sum_{j=1}^ns_j,l\right)\right\}
\end{align}
where
\begin{align}
&\Phi(k,i,l)={k-1-l \choose k-1-i}\frac{1}{(k-2)(k-1)^{k-1-l}}+{l \choose k-i}\frac{1}{(k-1)^l},  \nonumber\\
&\Psi(k,i,l)={k-1-l \choose k-i}\frac{1}{(k-1)^{k-1-l}}+{l \choose k-1-i}\frac{1}{(k-2)(k-1)^l}. \nonumber
\end{align}

Combining Eqs~(\ref{rhoA}-\ref{sigma_rule}) and normalizing $\sigma_{s_1s_2\cdots s_n}$ by dividing $(k^2(k+1)(k+2))$, we have
\begin{align}
\rho_A-\rho_B = \omega\sum_{s_1=0}^{g_1}\sum_{s_2=0}^{g_2}\cdots\sum_{s_n=0}^{g_n}\sigma_{s_1s_2\cdots s_n}\left(a_{s_1s_2\cdots s_n}-b_{(g_1-s_1)(g_2-s_2)\cdots (g_n-s_n)}\right)
\end{align}
where $\sigma_{s_1s_2\cdots s_n}$ is the normalized structure coefficients, given by
\begin{align} \label{sigma_value}
\sigma_{s_1s_2\cdots s_n}=
&\frac{(k-2)^{(k-\sum_{j=1}^ns_j)}}{k^2(k+1)(k+2)}\frac{\Pi_{j=1}^n{g_j \choose s_j}}
{{k \choose \sum_{j=1}^ns_j}}\nonumber \\
&\sum_{l=0}^{k}(k-l)\left\{\left[k^2-(k-2)l\right]\Phi\left(k,\sum_{j=1}^ns_j,l\right)+\left[2k+(k-2)l\right]\Psi\left(k,\sum_{j=1}^ns_j,l\right)\right\}.
\end{align}
This equation corresponds to Eq (1) in the main text.

\subsection{Replicator equation}
Infinite populations usually serve as a baseline model to investigate the evolutionary dynamics of a system.
Therefore we conduct a consistent investigation in infinite populations.
The evolutionary dynamics of multiplayer games on graphs with edge diversity can be described in terms of replicator equation.
Substituting Eqs~(\ref{cji}) and (\ref{dji}) into Eq~(\ref{mpA2}) and applying Eq~(\ref{identity_a3}), we have the replicator equation for evolutionary multiplayer games on graphs with $n$ types of edges
\begin{align}
\dot{x}=\frac{\omega(k-2)x(1-x)}{k^2}f(x) \label{replicator_equation1}
\end{align}
where
\begin{align}
f(x)=
&\sum_{s_1=0}^{g_1}\sum_{s_2=0}^{g_2}\cdots \sum_{s_n=0}^{g_n}
\left[\prod_{j=1}^n {g_j \choose s_j}x^{s_j}(1-x)^{g_j-s_j}\right]\left(\Lambda_a-\Lambda_b\right),
\label{replicator_equation}
\end{align}
\begin{align}
\Lambda_a=
&\sum_{r_1=0}^{g_1-s_1}\sum_{r_2=0}^{g_2-s_2}\cdots \sum_{r_n=0}^{g_n-s_n}
\left[\prod_{j=1}^n {g_j-s_j \choose r_j}z^{r_j}(1-z)^{g_j-s_j-r_j}\right]\nonumber \\
&\sum_{j=1}^n\left[\left(s_j+r_j\right)a_{(s_1+r_1)(s_2+r_2)\cdots(s_n+r_n)}
+\left(zs_j+\frac{r_j}{z}\right)a_{(s_1+r_1-\delta_{1j})(s_2+r_2-\delta_{2j})\cdots(s_n+r_n-\delta_{nj})}\right], \nonumber\\
\Lambda_b=
&\sum_{r_1=0}^{s_1}\sum_{r_2=0}^{s_2}\cdots \sum_{r_n=0}^{s_n}
\left[\prod_{j=1}^n {s_j \choose r_j}z^{r_j}(1-z)^{s_j-r_j}\right] \nonumber \\
&\sum_{j=1}^n\left[\left(g_j-s_j+r_j\right)b_{(s_1-r_1)(s_2-r_2)\cdots(s_n-r_n)}
+\left(z(g_j-s_j)+\frac{r_j}{z}\right)b_{(s_1-r_1+\delta_{1j})(s_2-r_2+\delta_{2j})\cdots(s_n-r_n+\delta_{nj})}\right]. \nonumber
\end{align}
This seemingly complicated Eq~(\ref{replicator_equation}) could be greatly simplified when applied to specific examples, such as traditional multiplayer games or pairwise games on graphs \cite{2006-Ohtsuki-p86-97}.

\clearpage
\section{Section 2. Recover the previous results as a specific case with $n=1$}
We can recover previous results in Ref. \cite{2016-Pena-p1-15} as a specific case by taking $g_1=k$ and $g_i=0$ for $i\ne1$, and thus rewrite the structure coefficient as
\begin{align}%\label{sigma_n1}
\sigma_s=
&\frac{(k-2)^{(k-s)}}{k^2(k+1)(k+2)}\sum_{l=0}^{k}(k-l)\left\{\left[k^2-(k-2)l\right]\Phi(k,s,l)+\left[2k+(k-2)l\right]\Psi(k,s,l)\right\}
\nonumber
\end{align}
where
\begin{align}
&\Phi(k,i,l)={k-1-l \choose k-1-i}\frac{1}{(k-2)(k-1)^{k-1-l}}+{l \choose k-i}\frac{1}{(k-1)^l},  \nonumber\\
&\Psi(k,i,l)={k-1-l \choose k-i}\frac{1}{(k-1)^{k-1-l}}+{l \choose k-1-i}\frac{1}{(k-2)(k-1)^l}. \nonumber
\end{align}

We can also recover the previous results by assuming that two $A-$players belonging to different types have an identical impact to their common opponent.
Plainly, $a_{s_1s_2\cdots s_n}$ and $b_{s_1s_2\cdots s_n}$ are unchanged if $\sum_{j}s_j$ is fixed.
Then the sum of $\sigma_{s_1s_2\cdots s_n}$ for all configurations satisfying $\sum_{j=1}^ns_j=s$ corresponds to the structure coefficient of term $a_{s}-b_{k-s}$.
Hence, we have
\begin{align}\label{sigma_n2}
\sigma_{s}=
&\sum_{\sum_{j=1}^n s_j=s}\sigma_{s_1s_2\cdots s_n} \nonumber\\
=&\sum_{\sum_{j=1}^n s_j=s}
\frac{(k-2)^{(k-1-s)}}{k^2(k+1)(k+2)}\frac{\Pi_{j=1}^n{g_j \choose s_j}}
{{k \choose \sum_{j=1}^ns_j}}\nonumber \\
&\sum_{l=0}^{k}(k-l)\left\{\left[k^2-(k-2)l\right]\Phi(k,s,l)+\left[2k+(k-2)l\right]\Psi(k,s,l)\right\}\nonumber \\
=&\frac{(k-2)^{(k-1-s)}}{k^2(k+1)(k+2)}\sum_{l=0}^{k}(k-l)\left\{\left[k^2-(k-2)l\right]\Phi(k,s,l)+\left[2k+(k-2)l\right]\Psi(k,s,l)\right\}.
\end{align}

\clearpage
\section{Section 3. Diverse multiplayer games}
The number of edges of type $i$ is $g_i$.
We designate $m$ the number of different values among all $g_i$s ($1\le i\le n$), $l_j$ ($1\le j\le m$) the $m$ corresponding values, and $n_j$ the number of edge types having $l_j$ edges.
Accordingly, we have
$\sum_{j=1}^m n_j=n$ and $\sum_{j=1}^m l_jn_j=k$.
For a clear description, we designate $g_{i}=l_j$ for $i\in V_j=\left[\sum_{u=1}^{j-1}n_u+1,\sum_{u=1}^{j}n_u\right]$.
Here we investigate a scenario where each individual plays different games with different individuals simultaneously.
We let individuals linked by the same type of edges form a group to play a multiplayer game.
Games defined in different types of edges are independent and thus could be different in both game metaphors or payoff entries.
Then the payoff can be reduced to
\begin{align}
&a_{s_1s_2\cdots s_n}=a_{s_1}^1+a_{s_2}^2+\cdots+a_{s_n}^n, \label{application_a} \\
&b_{s_1s_2\cdots s_n}=b_{s_1}^1+b_{s_2}^2+\cdots+b_{s_n}^n. \label{application_b}
\end{align}
$a_{s_i}^i$ ($b_{s_i}^i$) represents the payoff of an $A-$player (a $B-$player) obtained from the interaction with individuals of type $i$ where there are $s_i$ opposing $A$-players.

\subsection{Finite populations}
We first investigate how the independence of payoffs obtained in different games affects the ``sigma rule'' [see Eq~(\ref{sigma_rule})].
Using the above notations, we have
\begin{align}
&\sum_{s_1=0}^{g_1}\sum_{s_2=0}^{g_2}\cdots\sum_{s_n=0}^{g_n}\sigma_{s_1s_2\cdots s_n}\left(a_{s_1s_2\cdots s_n}-b_{(g_1-s_1)(g_2-s_2)\cdots (g_n-s_n)}\right)>0 \nonumber\\
\Longleftrightarrow
&\sum_{s_1=0}^{g_1}\sum_{s_2=0}^{g_2}\cdots\sum_{s_n=0}^{g_n}\sigma_{s_1s_2\cdots s_n}\sum_{i=1}^{n}(a_{s_i}^{i}-b_{g_i-s_i}^{i})>0 \nonumber \\
\Longleftrightarrow
&\sum_{s_1=0}^{g_1}\sum_{s_2=0}^{g_2}\cdots\sum_{s_n=0}^{g_n}\sigma_{s_1s_2\cdots s_n}\sum_{j=1}^{m}\sum_{i\in V_j}(a_{s_i}^{i}-b_{g_i-s_i}^{i})>0 \nonumber\\
\Longleftrightarrow
&\sum_{j=1}^{m}\sum_{i\in V_j}\sum_{s_1=0}^{g_1}\sum_{s_2=0}^{g_2}\cdots\sum_{s_n=0}^{g_n}\sigma_{s_1s_2\cdots s_n}(a_{s_i}^{i}-b_{g_i-s_i}^{i})>0 \label{diverse_label2}
\end{align}
Here we analyze the case for $j=1$, $V_1=\left[1,n_1\right]$, and $g_1=g_2=\cdots=g_{n_1}=l_1$.
Other cases can be calculated analogously.
We have
\begin{align}
&\sum_{i\in V_1}\sum_{s_1=0}^{g_1}\sum_{s_2=0}^{g_2}\cdots\sum_{s_n=0}^{g_n}\sigma_{s_1s_2\cdots s_n}(a_{s_i}^{i}-b_{g_i-s_i}^{i})
\nonumber \\
= &\sum_{s_1=0}^{g_1}(a_{s_1}^{1}-b_{g_1-s_1}^{1})\sum_{s_2=0}^{g_2}\sum_{s_3=0}^{g_3}\cdots\sum_{s_n=0}^{g_n}\sigma_{s_1s_2\cdots s_n}+\sum_{s_2=0}^{g_2}(a_{s_2}^{2}-b_{g_2-s_2}^{2})\sum_{s_1=0}^{g_1}\sum_{s_3=0}^{g_3}\cdots\sum_{s_n=0}^{g_n}\sigma_{s_1s_2\cdots s_n}+\cdots\nonumber\\
& +\sum_{s_{n_1}=0}^{g_{n_1}}(a_{s_{n_1}}^{n_1}-b_{g_{n_1}-s_{n_1}}^{n_1})\sum_{s_1=0}^{g_1}\cdots\sum_{s_{n_1-1}=0}^{g_{n_1-1}}
\sum_{s_{n_1+1}=0}^{g_{n_1+1}}\cdots\sum_{s_n=0}^{g_n}\sigma_{s_1s_2\cdots s_n} \label{diverse_label1}
\end{align}
From Eq~(\ref{sigma_n2}), we have
\begin{align}
\sigma_{s_1s_2\cdots s_n}=\frac{{k \choose \sum_{j=1}^ns_j}}
{\Pi_{j=1}^n{g_j \choose s_j}}\sigma_{\sum_{j=1}^ns_j}
\end{align}
Especially, for $g_i=g_j$, $\sigma_{s_1s_2\cdots s_n}$ remains unchanged after exchanging the $i_{th}$ and the $j_{th}$ subscripts of $\sigma_{s_1s_2\cdots s_n}$, i.e., $\sigma_{s_1\cdots s_{i-1}s_is_{i+1}\cdots s_{j-1}s_js_{j+1}\cdots s_n}=\sigma_{s_1\cdots s_{i-1}s_js_{i+1}\cdots s_{j-1}s_is_{j+1}\cdots s_n}$.
Denoting
\begin{align}
\sum_{s_2=0}^{g_2}\sum_{s_3=0}^{g_3}\cdots\sum_{s_n=0}^{g_n}\sigma_{s_1s_2\cdots s_n}=\tilde{\sigma}_{s_1}^1,
\end{align}
we have
\begin{align}
\sum_{s_1=0}^{g_1}\sum_{s_3=0}^{g_3}\cdots\sum_{s_n=0}^{g_n}\sigma_{s_1s_2\cdots s_n}=\sum_{s_1=0}^{g_1}\sum_{s_3=0}^{g_3}\cdots\sum_{s_n=0}^{g_n}\sigma_{s_2s_1\cdots s_n}=\tilde{\sigma}_{s_2}^1,
\end{align}
and
\begin{align}
\sum_{s_1=0}^{g_1}\cdots\sum_{s_{n_1-1}=0}^{g_{n_1-1}}
\sum_{s_{n_1+1}=0}^{g_{n_1+1}}\cdots\sum_{s_n=0}^{g_n}\sigma_{s_1s_2\cdots s_n}=\tilde{\sigma}_{s_{n_1}}^1.
\end{align}
Overall, Eq~(\ref{diverse_label1}) can be rewritten as
\begin{align} \label{diverse_label3}
&\sum_{s_1=0}^{g_1}(a_{s_1}^{1}-b_{g_1-s_1}^{1})\tilde{\sigma}_{s_1}^1+\sum_{s_2=0}^{g_2}(a_{s_2}^{2}-b_{g_2-s_2}^{2})\tilde{\sigma}_{s_2}^1
+\cdots+\sum_{s_{n_1}=0}^{g_{n_1}}(a_{s_{n_1}}^{n_1}-b_{g_{n_1}-s_{n_1}}^{n_1})\tilde{\sigma}_{s_{n_1}}^1 \nonumber\\
=&\sum_{s=0}^{l_1}\tilde{\sigma}_{s}^1\left(\sum_{i\in V_1}a_s^i-\sum_{i\in V_1}b_{l_1-s}^i\right)
\end{align}
Substituting Eq~\ref{diverse_label3} into Eq~\ref{diverse_label2}, we have
\begin{align}
\sum_{j=1}^{m}\sum_{s=0}^{l_j}\tilde{\sigma}_{s}^j\left(\sum_{i\in V_j}a_s^i-\sum_{i\in V_j}b_{l_j-s}^i\right)>0
\end{align}
where
\begin{align}
\tilde{\sigma}_{s}^j=\sum_{s_1=0}^{g_1}\cdots\sum_{s_{r-1}=0}^{g_{r-1}}\sum_{s_{r+1}=0}^{g_{r+1}}\cdots\sum_{s_n=0}^{g_n}\sigma_{s_1s_2\cdots s_n}
\end{align}
and $r=\sum_{u=1}^{j-1}n_u+1$.
Thus the effects of the population structure are captured by $\sum_{j=1}^m (l_j+1)$ structure coefficients.
Especially, for $g_1=g_2=\cdots=g_n=g$, $\rho_A>\rho_B$ is equivalent to
\begin{align}
\sum_{s=0}^{g}\tilde{\sigma}_{s}^1\left(\sum_{i=1}^n a_s^i-\sum_{i=1}^n b_{g-s}^i\right)>0.
\end{align}
Using Eqs~(\ref{rhoA}), (\ref{rhoB}), (\ref{alpha_final}) and (\ref{beta2}), we find that for sufficient large populations the fixation probabilities (both $\rho_A$ and $\rho_B$) under diverse multiplayer games can be approximated by assuming players playing a unified game, where the payoff structure correspond to the average over all games.

\subsection{Infinite populations}
We proceed with the study of diverse multiplayer games in infinite populations.
Applying Eqs~(\ref{application_a}) and (\ref{application_b}) into Eq~(\ref{replicator_equation}), we have
\begin{align}
f(x)=
&\sum_{j=1}^n\sum_{s_j=0}^{g_j}{g_j \choose s_j}x^{s_j}(1-x)^{g_j-s_j}\sum_{r_j=0}^{g_j-s_j}{g_j-s_j \choose r_j}z^{r_j}(1-z)^{g_j-s_j-r_j} \nonumber \\
&\left[\left[(1+z)(k-g_j)+s_j+r_j\right]a_{s_j+r_j}^j+\left(zs_j+\frac{r_j}{z}\right)a_{s_j+r_j-1}^j\right] \nonumber \\
&-\sum_{j=1}^n\sum_{s_j=0}^{g_j}{g_j \choose s_j}x^{s_j}(1-x)^{g_j-s_j}\sum_{r_j=0}^{s_j}{s_j \choose r_j}z^{r_j}(1-z)^{s_j-r_j} \nonumber \\
&\left[\left[(1+z)(k-g_j)+g_j-s_j+r_j\right]b_{s_j-r_j}^j+\left(z(g_j-s_j)+\frac{r_j}{z}\right)b_{s_j-r_j+1}^j\right] \nonumber\\
=&\sum_{i=1}^m\sum_{s=0}^{l_i}{l_i \choose s}x^{s}(1-x)^{l_i-s}\sum_{r=0}^{l_i-s}{l_i-s \choose r}z^{r}(1-z)^{l_i-s-r} \nonumber \\
&\left[\left[(1+z)(k-l_i)+s+r\right]\sum_{j\in V_i}a_{s+r}^j+\left(zs+\frac{r}{z}\right)\sum_{j\in V_i}a_{s+r-1}^j\right] \nonumber \\
&-\sum_{i=1}^m\sum_{s=0}^{l_i}{l_i \choose s}x^{s}(1-x)^{l_i-s}\sum_{r=0}^{s}{s \choose r}z^{r}(1-z)^{s-r} \nonumber \\
&\left[\left[(1+z)(k-l_i)+l_i-s+r\right]\sum_{j\in V_i}b_{s-r}^j+\left(z(l_i-s)+\frac{r}{z}\right)\sum_{j\in V_i}b_{s-r+1}^j\right].
\label{application_fx}
\end{align}
Furthermore, if we introduce two notations
\begin{align}
&\bar{a}_{s}^i=\frac{1}{n_i}\sum_{j\in V_i}a_{s}^j, \nonumber\\
&\bar{b}_{s}^i=\frac{1}{n_i}\sum_{j\in V_i}b_{s}^j, \nonumber
\end{align}
which correspond to the average of payoff values over games of same sizes, i.e., $g_j$s are identical for $j\in V_i$.
In other words, for games of same sizes, we can use the average of their payoff values to approximate the evolutionary dynamics.
Especially, for $g_1=g_2=\cdots=g_n=g$, the evolutionary dynamics can be approximated by a unified payoff structure
\begin{align}
&\bar{a}_{s}=\frac{1}{n}\sum_{j=1}^n a_{s}^j, \nonumber\\
&\bar{b}_{s}=\frac{1}{n}\sum_{j=1}^n b_{s}^j. \nonumber
\end{align}

We end this section by an example of evolutionary games on weighted networks with $g_1=g_2=\cdots=g_n=g$.
We endow the $j_{th}$ type of edges a weight $\zeta_j$.
The payoff structure is
\begin{align}
&a_{s}^j=\zeta_{j} a_{s}, \label{weighted_a} \\
&b_{s}^j=\zeta_{j} b_{s}, \label{weighted_b}
\end{align}
where $a_s$ ($b_s$) is a function of $s$.
Substituting Eqs~(\ref{weighted_a}) and (\ref{weighted_b}) into Eq~(\ref{application_fx}), we have that
\begin{align}
f(x)=
&\sum_{j=1}^n\zeta_j\cdot\sum_{s=0}^{g}{g \choose s}x^{s}(1-x)^{g-s} \nonumber\\
&\Bigg\{\sum_{r=0}^{g-s}{g-s \choose r}z^{r}(1-z)^{g-s-r}\left[\left[(1+z)(k-g)+s+r\right]a_{s+r}+\left(zs+\frac{r}{z}\right)a_{s+r-1}\right] \nonumber \\
&-\sum_{r=0}^{s}{s \choose r}z^{r}(1-z)^{s-r}\left[\left[(1+z)(k-g)+g-s+r\right]b_{s-r}+\left(z(g-s)+\frac{r}{z}\right)b_{s-r+1}\right]\Bigg\} \label{weighted_c}.
\end{align}
Equation (\ref{weighted_c}) shows that the values of $x$ satisfying $f(x)=0$ are independent of $\zeta_j$ for any $j$.
Thus, nonuniform strength of interactions does not affect the evolutionary dynamics.

\clearpage
\section{Section 4. Sigma rule and structure coefficient for evolutionary two-player games on graphs with $n$ types of edges}
In evolutionary two-player games on graphs, interactions occurring in each type of edges are assigned a payoff matrix.
The payoff matrix for interactions occurring in edges of type $i$ is
$$\bordermatrix{
  & \text{A} & \text{B} \cr
\text{A} & \alpha_i & \beta_i \cr
\text{B} & \gamma_i & \theta_i \cr
}$$
where each value corresponds to the payoff assigned to the individual adopting a strategy in the row against its partner taking a strategy in the column.
Transforming the payoff to multiplayer interactions through
$a_{s_1s_2\cdots s_n}=\sum_{i=1}^n \left[s_i\alpha_i+(g_i-s_i)\beta_i\right]$ and
$b_{s_1s_2\cdots s_n}=\sum_{i=1}^n \left[s_i\gamma_i+(g_i-s_i)\theta_i\right]$, we have the sigma rule from Eq~(\ref{sigma})
\begin{align}
\sum_{i=1}^n \bar{s}_i\alpha_i+\sum_{i=1}^n \left(g_i-\bar{s}_i\right)\beta_i
-\sum_{i=1}^n \left(g_i-\bar{s}_i\right)\gamma_i-\sum_{i=1}^n \bar{s}_i\theta_i>0 \label{sigma_pairwise}
\end{align}
where
\begin{align}
\bar{s}_i = \sum_{s_1=0}^{g_1}\sum_{s_2=0}^{g_2}\cdots\sum_{s_n=0}^{g_n}\sigma_{s_1s_2\cdots s_n}s_i \nonumber.
\end{align}
Here we show how to get $\bar{s}_i$ relying on a previous study \cite{2007-Ohtsuki-p108106-108106}.
Assuming that interactions along all edges except edges of type $i$ bring no benefits ($\alpha_j=0,\beta_j=0,\gamma_j=0,\theta_j=0$ for $j\ne i$),
Eq~\ref{sigma_pairwise} can be rewritten as
\begin{align}
\rho_A>\rho_B \Longleftrightarrow \bar{s}_i\alpha_i+ \left(g_i-\bar{s}_i\right)\beta_i
- \left(g_i-\bar{s}_i\right)\gamma_i- \bar{s}_i\theta_i>0. \label{sigma_rule_simplified}
\end{align}
From the perspective of separated interaction graph and replacement graph, in the interaction graph, these edges corresponding to interactions with no payoffs seem to be removed, leading to asymmetric interaction and replacement graphs.
From Ref.~\cite{2007-Ohtsuki-p108106-108106}, we have
\begin{align}
\rho_A>\rho_B \Longleftrightarrow (k+1)\alpha_i+ (k-1)\beta_i
- (k-1)\gamma_i- (k+1)\theta_i>0 . \label{sigma_rule_Ohtsuki}
\end{align}
Comparing Eqs (\ref{sigma_rule_simplified}) and (\ref{sigma_rule_Ohtsuki}), we have
\begin{align}
\bar{s}_i = \frac{g_i(k+1)}{2k}.
\end{align}
Thus we have sigma rule shown in the main text
\begin{align}
\sum_{i=1}^n\left[g_i(k+1)\alpha_i+g_i(k-1)\beta_i\right]>\sum_{i=1}^n\left[g_i(k-1)\gamma_i+g_i(k+1)\theta_i\right] \nonumber.
\end{align}

\nolinenumbers

\end{document}